\documentstyle[prb,aps,epsfig,a4]{revtex}
\begin{document}        

\title{\bf Study of the process $e^+e^- \to \pi^+\pi^-\pi^0$ in the energy
           region $\sqrt[]{s}$ from 0.98 to 1.38 GeV.}
\author{ M.N.Achasov\thanks{e-mail: achasov@inp.nsk.su, 
FAX: +7(383-2)34-21-63},
         V.M.Aulchenko, K.I.Beloborodov, A.V.Berdyugin,A.G.Bogdanchikov,
	 A.V.Bozhenok, A.D.Bukin, D.A.Bukin, S.V.Burdin, T.V.Dimova,
	 V.P.Druzhinin, V.B.Golubev, V.N.Ivanchenko, A.A.Korol,
         S.V.Koshuba, I.N.Nesterenko, E.V.Pakhtusova, A.A.Polunin,
	 A.A.Salnikov, S.I.Serednyakov,  V.V.Shary,
	 Yu.M.Shatunov, V.A.Sidorov, Z.K.Silagadze, A.N.Skrinsky,
	 A.G.Skripkin, Yu.V.Usov, A.V.Vasiljev} 
\address{Budker Institute of Nuclear Physics,  \\
         Siberian Branch of the Russian Academy of Sciences \\
         and Novosibirsk State University, \\
	 11 Lavrentyev,Novosibirsk, \\
	 630090, Russia} 
\date{\today}
\maketitle

\begin{abstract}
 The cross section of the process $e^+e^-\to \pi^+\pi^-\pi^0$ was measured in
 the Spherical Neutral Detector experiment at the VEPP-2M collider in the
 energy region $\sqrt[]{s} = 980 \div 1380$ MeV. The measured cross section,
 together with the $e^+e^-\to \pi^+\pi^-\pi^0$ and $\omega\pi^+\pi^-$ cross  
 sections obtained in other experiments,
 was analyzed in the framework of the generalized vector meson dominance model.
 It was found that the experimental
 data can be described by a sum of $\omega$, $\phi$ mesons and two
 $\omega^\prime$ and $\omega^{\prime\prime}$ resonances contributions, with
 masses $m_{\omega^\prime}\sim 1490$,
 $m_{\omega^{\prime\prime}}\sim 1790$ MeV and widths
 $\Gamma_{\omega^\prime}\sim 1210$,
 $\Gamma_{\omega^{\prime\prime}}\sim 560$ MeV. The analysis of the
 $\pi^+\pi^-$ invariant mass spectra in the energy region $\sqrt[]{s}$ 
 from 1100 to 1380 MeV has shown that for their description 
 one should take into account  the $e^+e^-\to\omega\pi^0\to\pi^+\pi^-\pi^0$ 
 mechanism also. The
 phase between the amplitudes corresponding to the 
 $e^+e^-\to\omega\pi$ and $e^+e^-\to\rho\pi$ intermediate states
 was measured for the first time. The value of the phase is close to zero 
 and depends on energy.
\end{abstract}

\section{Introduction}

 The cross section of hadron production in the $e^+e^-$ annihilation
 in the energy region $\sqrt[]{s} < 1.03$ GeV can be described within the
 vector meson dominance model (VDM) framework 
 and is determined by the transitions of light vector mesons
 ($\rho,\omega,\phi$) into the final states.
 The light vector mesons have been studied rather well. They are
 quark-antiquark
 $q\overline{q}$ ($q=u,d,s$) bound states, and their masses, widths and  
 main decays have been
 measured with high accuracy \cite{pdg}. The cross section
 for hadron production above the $\phi(1020)$ resonance
 ($\sqrt[]{s} \simeq 1.03$--$2$ GeV) cannot be described in the conventional
 VDM framework (taking into account $\rho,\omega$ and $\phi$ mesons only)
 indicating the existence of states with vector
 meson quantum numbers $I^G(J^{PC})=1^+(1^{--}),0^-(1^{--})$ and with masses of
 about 1450, 1650 MeV. Parameters of these states are not well established due
 to inaccurate and conflicting experimental data. The nature of
 these states is not clear either. In some reviews of experimental data they 
 are considered as a mixture of $q\overline{q}$ with 4-quark $qq\overline{qq}$
 and hybrid
 $q\overline{q}g$ states \cite{don1,don2,don3,don4}. On the other hand, the
 experimental data do not contradict the hypothesis that these states have
 $q\overline{q}$ structure and are radial and orbital excitations of the
 light vector mesons \cite{ak1,ak2,ak3}. In this
 context the main experimental task is the improvement of the accuracy of 
 cross section measurement.

 As was already mentioned, in the VDM framework the cross section of the 
 process $e^+e^- \to \pi^+\pi^-\pi^0$ is determined  by the amplitudes 
 of vector meson
 $V$ ($V=\omega,\phi,\omega^\prime,{\ldots}$) transitions into the final state:
 $V \to\pi^+\pi^-\pi^0$. The $\rho\pi$ intermediate state dominates in these
 transitions [Fig.\ref{diag}(a)]. The other mechanism of $V\to\pi^+\pi^-\pi^0$
 transition is also possible via $\rho-\omega$ mixing:
 $V\to\omega\pi^0\to\rho^0\pi^0$ ($V=\rho,\rho^\prime,\rho^{\prime\prime}$)
 [Fig.\ref{diag}(b)]. This effect was predicted in Ref.\cite{thrhoom}
 and was observed in the SND (Spherical Neutral Detector) experiment in the
 energy range $\sqrt[]{s}=1200$--$1400$ MeV \cite{sndro}. The studies 
 of the
 $e^+e^- \to \pi^+\pi^-\pi^0$ cross section and $\pi\pi$ invariant mass spectra
 above $\phi$-meson production region provide information about excited states
 of vector meson and their interference.
 
 The $e^+e^- \to \pi^+\pi^-\pi^0$ cross section in the energy region above 
 $\phi$ meson and up to 2200 MeV has been studied in several experiments
 \cite{m3n,mea,gg2,dm1,nd,dm2}, but none of them have covered the whole
  region.
 The SND study of this cross section in the range $\sqrt[]{s} = 1040$ -- 1380
 MeV based on a part of collected data was already reported in
 Ref.\cite{sndmhad}. Here we present the results obtained by using the 
 total data
 sample. The present work includes both the total cross section and the
 dipion mass spectra studies.

\section{Experiment}

 The SND detector \cite{sndnim} ran from 1995 to 2000 at the
 VEPP-2M \cite{vepp2} collider in the energy range $\sqrt[]{s}$ from 360 to
 1400 MeV. The detector contains several subsystems. The tracking system
 includes two cylindrical drift chambers. The three-layer spherical
 electromagnetic calorimeter is based on NaI(Tl) crystals \cite{calor99}.
 The muon/veto
 system consists of plastic scintillation counters and two layers of streamer
 tubes. The calorimeter energy and angular resolution depends on the photon
 energy as $\sigma_E/E(\%) = {4.2\% / \sqrt[4]{E(\mbox{GeV})}}$ and
 $\sigma_{\phi,\theta} = {0.82^\circ / \sqrt[]{E(\mathrm{GeV})}} \oplus 0.63^\circ$
 The tracking system angular resolution is about $0.5^\circ$ and $2^\circ$ for
 azimuthal and polar angles respectively. The energy loss resolution $dE/dx$ in
 the drift chamber is about 30\%. SND was described in details in
 Ref.\cite{sndnim}.

 In 1997 and 1999 the SND collected data in the energy region $\sqrt[]{s}$
 from 1040 to 1380 MeV with integrated luminosity about $9.0~\mbox{pb}^{-1}$,
 in addition about $130~\mbox{nb}^{-1}$ was collected at $\sqrt[]{s}=980$ MeV.
 The beam energy was calculated from the magnetic field value in the bending
 magnets and revolution frequency of the collider. The center of mass energy
 determination accuracy is about 0.1 MeV and the spread of the beam energy is 
 from 0.2 to 0.4 MeV.

 For the luminosity measurements, the processes $e^+e^- \to e^+e^-$ and
 $e^+e^- \to \gamma\gamma$ were used. In this work the luminosity measured by
 $e^+e^- \to e^+e^-$ was used for normalization.
 The systematic error of the integrated luminosity determination is estimated
 to be 2\%. Since luminosity measurements by  $e^+e^- \to e^+e^-$ and
 $e^+e^- \to \gamma\gamma$ reveal a systematic spread of about 1\%, this  was
 added to the statistical error of the luminosity determination in each
 energy point. The statistical accuracy was better than 1\%.

\section{Data analysis}

\subsection{Selection of $e^+e^- \to \pi^+\pi^-\pi^0$ events}

 The data analysis and selection criteria used in this work are similar
 to those described in Ref.\cite{phi98,dplphi98}. During the experimental runs,
 the first-level trigger \cite{sndnim} selects events with energy deposition
 in the calorimeter more than 180 MeV and with two or more charged particles.
 During processing of the experimental data the event reconstruction is
 performed \cite{sndnim,phi98}. For further analysis, events containing two or
 more photons and two charged particles with $|z| < 10$ cm and $r < 1$ cm were
 selected. Here $z$ is the coordinate of the charged particle production point
 along the beam axis (the longitudinal size of the interaction region depends
 on beam energy and varies from 2 to 2.5 cm); $r$ is the distance between the
 charged particle track and the beam axis in the $r-\phi$ plane. Extra photons
 in $e^+e^- \to \pi^+\pi^-\pi^0$ events can appear because of the overlap with
 the beam background or nuclear interactions of the charged pions in the
 calorimeter. Under these selection conditions, the background sources
 are  $e^+e^- \to \pi^+\pi^-\pi^0\pi^0$, $e^+e^-\gamma\gamma$,
 $\pi^+\pi^-\gamma$, $K^+K^-$, $K_SK_L$ processes and the beam background. We
 note that in the energy region above the $\phi$-meson the process
 $e^+e^- \to \pi^+\pi^-\pi^0$ does not dominate. Even more, its cross section
 is several times lower than the cross section of the main background process
 $e^+e^- \to \pi^+\pi^-\pi^0\pi^0$.

 To suppress the beam background, the following cuts on the angle $\psi$ 
 between two charged particle tracks and energy deposition of the neutral
 particles $E_{neu}$ were applied: $\psi > 40^{\circ}$, $E_{neu}>100$ MeV.

 To reject the background from the $e^+e^- \to K^+K^-$ process, the following
 cuts were imposed: $(dE/dx) < 5 \cdot (dE/dx)_{min} $ for each charged
 particle, $(dE/dx) < 3 \cdot (dE/dx)_{min} $ at least for one of them, and
 $\Delta\phi > 10^\circ$. Here $\Delta\phi$ is an acollinearity angle in the
 azimuthal plane and $(dE/dx)_{min}$ is an average energy loss of a minimum
 ionizing particle.  The last cut $|\Delta\phi| > 10^\circ$ also suppresses
 the  $e^+e^- \to \pi^+\pi^-\gamma$ events.
 
 To suppress the  $e^+e^- \to e^+e^-\gamma\gamma$ events an
 energy deposition in the calorimeter of the charged particles $E_{cha}$ was
 required to be small enough: $E_{cha} < 0.5 \cdot \sqrt[]{s}$.

 For events left after these cuts, a kinematic fit was performed under the
 following constraints: the charged particles are assumed to be pions, the
 system has zero total momentum, the total energy is $\sqrt[]{s}$, and the
 photons originate from the $\pi^0 \to \gamma\gamma$ decays. The value of the
 $\chi^2$ function $\chi^2_{3\pi}$ (Fig.\ref{xi2u}) is calculated during the
 fit. In events with more than two photons, extra photons are considered as
 spurious ones and rejected. To do this, all possible subsets of two photons
 were inspected and the one, corresponding to the maximum likelihood was
 selected. After the kinematic fit the following additional cuts were applied:
 $N_{\gamma}=2$ ($N_{\gamma}$ is the number of detected photons),
 $\chi^2_{3\pi} < 5$ and the polar angle $\theta_\gamma$ of at least one of
 the photons should satisfy to the following criterion:
 $36^{\circ} < \theta_\gamma < 144^{\circ}$.
 The angular distributions of particles for the selected events are shown in
 Fig.\ref{tetn},\ref{teu12},\ref{teu3} and \ref{teu45} while Fig.\ref{epu4}
 and Fig.\ref{epu45} demonstrate the photon energy distributions for the same
 events. The experimental and simulated distributions are in agreement.

\subsection{Background subtraction}

 The number of background events was estimated from the following formula:
\begin{eqnarray}
\label{bg}
 N_{bkg}({s}) = \sum_i \sigma_{Ri}({s}) \epsilon_i({s})  IL({s}),
\end{eqnarray}
 where $i$ is a process number, $\sigma_{Ri}({s})$ is the cross
 section of the background process taking into account the radiative
 corrections, $IL({s})$ is the integrated luminosity, $\epsilon_i({s})$ is
 the detection probability for the background process obtained from simulation
 under selection described above. The $e^+e^-\to\pi^+\pi^-\gamma$ cross
 section was calculated for the case, when the photon has the energy above 10
 MeV and is radiated at the angle $\theta$ more than $10^\circ$. As it was
 mentioned above, the main source of background is the events of the
 $e^+e^-\to\pi^+\pi^-\pi^0\pi^0$ process. Two mechanisms contribute to
 the total cross section of this process:
 $e^+e^-\to\omega\pi$ and $e^+e^-\to\rho\pi\pi$. It was shown in
 Ref.\cite{cmd4p,cleo4p} that the $e^+e^-\to\rho\pi\pi$ process dynamics can
 be described with the $a_1\pi$ intermediate state. The SND studies of the
 $e^+e^-\to\pi^+\pi^-\pi^0\pi^0$ process \cite{snd4p} agree with this
 conclusion. For background estimation the $e^+e^-\to\omega\pi$
 and $e^+e^-\to\rho\pi\pi$
 cross sections  measured in SND experiments were used
 \cite{snd4p,ppg}. To obtain the detection probability of the
 $e^+e^-\to\rho\pi\pi$ events, the simulation with the $a_1\pi$ intermediate
 state was used. The numbers of $e^+e^-\to \pi^+\pi^-\pi^0(\gamma)$ events
 (after background subtraction) and background event numbers are shown in
 Table~\ref{tab1}. Here $\gamma$ is a photon emitted by initial particles.

 To estimate the accuracy of background events number determination the
 $\chi^2_{3\pi}$ distribution (Fig.\ref{xi2u}) was studied. The experimental
 $\chi^2_{3\pi}$ distribution in the range $0<\chi^2_{3\pi}<20$ was fitted
 by a sum of background and signal. The distribution for background events
 was taken from the simulation and that for $e^+e^-\to\pi^+\pi^-\pi^0$
 events was obtained by using data collected in the vicinity of the $\phi$
 meson peak \cite{phi98,dplphi98} (the $\chi^2_{3\pi}$ distribution actually 
 does not change in the interval $\sqrt[]{s}= 1$ -- 1.4 GeV). 
 As a result, the ratio between the number of background events obtained
 from the fit and the number calculated according to (\ref{bg}) was found
 to be $1.4 \pm 0.2$. Using this ratio, the accuracy of the determination 
 of the number of background events can be estimated to be about 40\%.

\subsection{Detection efficiency}

 The detection efficiency of the $e^+e^-\to\pi^+\pi^-\pi^0(\gamma)$
 process was obtained from simulation. The detection efficiency for
 events without $\gamma$-quantum radiation depends on the center of mass energy
 and varies from 0.15 to 0.16 in the energy range 
 $\sqrt[]{s}=980$ -- 1380 MeV. This
 dependence can be approximated by a linear function. The detection efficiency
 dependence on the radiated photon energy is shown in Fig.\ref{efrad} .

 Inaccuracies in the simulation of the
 $\chi^2_{3\pi}$, $dE/dx$, and $N_\gamma$ distributions lead to an error in the 
 average detection efficiency determination. To take into account these 
 uncertainties, the detection efficiency was multiplied by correction
 coefficients, which were obtained in the following way \cite{phi98}. The
 experimental events were selected without any conditions on the parameter
 under study, using the selection parameters uncorrelated with the studied one.
 The same selection criteria were applied to simulated events. Then the cut was
 applied to the parameter and the correction coefficient was calculated:
\begin{eqnarray}
 \delta = { {n/N} \over {m/M} },
\end{eqnarray}
 where $N$ and $M$ are the number of events in experiment and simulation
 respectively selected without any cuts on the parameter under study; 
 $n$ and $m$ are
 the number of events in experiment and simulation when the cut on the
 parameter was applied. As a rule, the error in the coefficient  $\delta$
 determination is connected with the uncertainty of background subtraction.
 This systematic error was estimated by varying other selection criteria.
 The correction coefficient $\delta_{\chi^2_{3\pi}}=0.91\pm0.03$, due to the 
 uncertainty in the $\chi^2_{3\pi}$ distribution simulation,
 was obtained using data collected in the
 vicinity of the $\phi$ resonance \cite{phi98,dplphi98}. The correction which
 takes into account the inaccuracy of simulation of extra photons is
 $\delta_{N_\gamma} = 0.87 \pm 0.02$, and that correction for the inaccuracy of
 simulation $dE/dx$ energy losses  is $\delta_{dE/dx} = 0.98 \pm 0.01$.
 The overlap of the beam background with the events containing charged
 particles can result in track reconstruction failure and a decrease of
 detection efficiency. To take into account this effect, background events
 (experimental events collected when the detector was triggered with an
 external generator) were mixed with the simulated events. It was found that the
 detection efficiency decreased by about 3\% and therefore the correction
 coefficient $\delta_{over} = 0.97 \pm 0.03$ was used.

 The total correction used in this work is equal to:
$$\delta_{tot}=\delta_{\chi^2_{3\pi}}\times\delta_{dE/dx}\times\delta_{N_\gamma}
\times\delta_{over} = 0.75 \pm 0.04.$$
 The systematic error of detection efficiency determination is 5\%. The
 detection efficiency after the applied corrections  is shown in 
 Table~\ref{tab1}.

\section{Theoretical framework}

 In the VDM framework the cross section of the  $e^+e^-\to\pi^+\pi^-\pi^0$
 process is
\begin{eqnarray}
\label{ds}
 {d\sigma \over dm_0 dm_+} = { {4\pi\alpha} \over {s^{3/2}} }
 {{|\vec{p}_+ \times \vec{p}_-|^2} \over {12\pi^2\mbox{~}\sqrt[]{s}}} m_0m_+ \cdot
 |F|^2,
\end{eqnarray}
 where $\vec{p}_+$ and $\vec{p}_-$ are the $\pi^+$ and $\pi^-$
 momenta, $m_0$ and $m_+$ are $\pi^+\pi^-$ and $\pi^+\pi^0$ invariant masses.
 The formfactor $F$ of the $\gamma^\star \to \pi^+\pi^-\pi^0$ transition
 has the form
\begin{eqnarray}
\label{formfac}
 |F|^2 = \Biggl| A_{\rho\pi}(s)  \sum_{i=+,0,-} 
 { g_{\rho^i\pi\pi} \over D_\rho(m_i)} +
 A_{\omega\pi}(s)
 {\Pi_{\rho\omega}g_{\rho^0\pi\pi}\over D_\rho(m_0) D_\omega(m_0)} \Biggr|^2 .
\end{eqnarray}
 Here
$$D_\rho(m_i) = m_{\rho^i}^2 - m_i^2 -im_i\Gamma_{\rho^i}(m_i),$$
$$\Gamma_{\rho^i}(m_i) = \Biggl({m_{\rho^i} \over m_i}\Biggr)^2 \cdot
  \Gamma_{\rho^i} \cdot  \Biggl({q_i(m_i) \over q_i(m_{\rho^i})}\Biggr)^3$$
$$q_0(m) = {1 \over 2}(m^2-4m_\pi^2)^{1/2},$$
$$q_\pm(m) = {1 \over 2m}
 \bigl[(m^2-(m_{\pi^0}+m_\pi)^2)(m^2-(m_{\pi^0}-m_\pi)^2)\bigr]^{1/2}$$
 $$m_-=\sqrt[]{s+m_{\pi^0}^2+2m_{\pi}^2-m_0^2-m_+^2},$$
 where $m_-$ is the $\pi^-\pi^0$ invariant mass, $m_{\pi^0}$ and $m_\pi$ are
 the neutral and charged pion masses, $i$ denotes the sign of a $\rho$-meson
 ($\pi\pi$ pair) charge. The $\rho^0 \to \pi^+\pi^-$ and
 $\rho^\pm\to\pi^\pm\pi^0$ transition coupling constants could be determined
 in the following way:
 $$g_{\rho^0\pi\pi}^2 = {6\pi m_{\rho^0}^2\Gamma_{\rho^0} \over
 q_0(m_{\rho^0})^3},$$
$$ g_{\rho^\pm\pi\pi}^2 = {6\pi m_{\rho^\pm}^2\Gamma_{\rho^\pm} \over
 q_\pm(m_{\rho^\pm})^3}$$
 Experimental data \cite{dplphi98} do not contradict the equality 
 of the coupling constants
 $g_{\rho^0\pi\pi}^2 = g_{\rho^\pm\pi\pi}^2$. In this case the $\rho^0$ and
 $\rho^\pm$ meson widths are related as follows:
\begin{eqnarray}
\label{shir}
 \Gamma_{\rho^\pm} = \Gamma_{\rho^0}{m_{\rho^0}^2 \over m_{\rho^\pm}^2}
 { q_\pm(m_{\rho^\pm})^3 \over q_0(m_{\rho^0})^3}.
\end{eqnarray}
 In the subsequent analysis we assume that
 $g_{\rho^0\pi\pi}^2 = g_{\rho^\pm\pi\pi}^2$, and the width values were
 taken from SND measurements \cite{dplphi98} $\Gamma_{\rho^0} = 149.8$ MeV,
 $\Gamma_{\rho^\pm} = 150.9$ MeV. The neutral and charged $\rho$ mesons masses
 were assumed to be equal and were also taken from the SND measurements 
 \cite{dplphi98}
 $m_\rho=775.0$ MeV.
 
 The second term in (\ref{formfac}) takes into account the  $\rho-\omega$ 
 mixing \cite{thrhoom}. The polarization operator of this mixing 
 $\Pi_{\rho\omega}$ satisfies
 $\mbox{Im}(\Pi_{\rho\omega}) \ll \mbox{Re}(\Pi_{\rho\omega})$ 
 \cite{akfaz,akozi}, where
\begin{eqnarray}
 \mbox{Re}(\Pi_{\rho\omega}) =
 \sqrt[]{{\Gamma_\omega \over \Gamma_{\rho^0}(m_\omega)} 
 B(\omega\to\pi^+\pi^-)} \cdot \biggl| (m_\omega^2-m_\rho^2) - 
 im_\omega(\Gamma_\omega - \Gamma_{\rho^0}(m_\omega))\biggr|,
\end{eqnarray}
 so we assumed $\mbox{Im}(\Pi_{\rho\omega}) = 0$ in the subsequent 
 analysis.

 The $e^+e^-\to\pi^+\pi^-\pi^0$ process cross section can be written in the
 following way:
\begin{eqnarray}
\label{sech3p}
 \sigma_{3\pi} = \sigma_{\rho\pi\to3\pi} + \sigma_{\omega\pi\to3\pi} +
 \sigma_{int},
\end{eqnarray}
 where
\begin{eqnarray}
\label{sech1}
 \sigma_{\rho\pi\to3\pi} = {{4\pi\alpha} \over {s^{3/2}}}
 W_{\rho\pi}(s)\biggl| A_{\rho\pi}(s) \biggr|^2,
\end{eqnarray}
\begin{eqnarray}
\label{sech2}
 \sigma_{\omega\pi\to3\pi} = {{4\pi\alpha} \over {s^{3/2}}}
 W_{\omega\pi}(s)\biggl| A_{\omega\pi}(s) \biggr|^2,
\end{eqnarray}
\begin{eqnarray}
\label{sech3}
 \sigma_{int} = {{4\pi\alpha} \over {s^{3/2}}}
 \biggl\{ A_{\rho\pi}(s) A_{\omega\pi}^\star (s) W_{int}(s) +
 A_{\rho\pi}^\star (s) A_{\omega\pi}(s) W_{int}^\star (s) \biggr\}.
\end{eqnarray}
 The phase space factors $W_{\rho\pi}(s)$, $W_{\omega\pi}(s)$ and $W_{int}(s)$
 were calculated as follows:
\begin{eqnarray}
 W_{\rho\pi}(s) = {1 \over 12 \pi^2 \mbox{~}\sqrt[]{s}}
 \int\limits^{\sqrt[]{s}-m_{\pi^0}}_{2m_\pi}
 m_0 dm_0 \int\limits^{m_+^{max}(m_0)}_{m_+^{min}(m_0)}
 m_+ dm_+ |\vec{p}_+ \times \vec{p}_-|^2 \cdot \biggl|\sum_{i=+,0,-} 
 { g_{\rho^i\pi\pi} \over D_\rho(m_i)}\biggr|^2,
\end{eqnarray}
\begin{eqnarray}
 W_{\omega\pi}(s) = {1 \over 12 \pi^2 \mbox{~}\sqrt[]{s}}
 \int\limits^{\sqrt[]{s}-m_{\pi^0}}_{2m_\pi}
 m_0 dm_0 \int\limits^{m_+^{max}(m_0)}_{m_+^{min}(m_0)}
 m_+ dm_+ |\vec{p}_+ \times \vec{p}_-|^2 \cdot
 \biggl| {\Pi_{\rho\omega}g_{\rho^0\pi\pi}\over D_\rho(m_0) D_\omega(m_0)}
 \biggr|^2,
\end{eqnarray}
\begin{eqnarray}
 W_{int}(s) = {1 \over 12 \pi^2 \mbox{~}\sqrt[]{s}}
 \int\limits^{\sqrt[]{s}-m_{\pi^0}}_{2m_\pi}
 m_0 dm_0 \int\limits^{m_+^{max}(m_0)}_{m_+^{min}(m_0)}
 m_+ dm_+ |\vec{p}_+ \times \vec{p}_-|^2 \cdot
 \biggl( \biggl[ {\Pi_{\rho\omega}g_{\rho^0\pi\pi}\over 
 D_\rho(m_0) D_\omega(m_0)}\biggr]^\star \cdot
 \sum_{i=+,0,-}{ g_{\rho^i\pi\pi}\over D_\rho(m_i)}
 \biggr),
\end{eqnarray}
  
 Amplitudes of the $\gamma^\star \to \rho\pi$ and
 $\gamma^\star \to \omega\pi^0$ transitions have the form
\begin{eqnarray}
\label{aropi}
 A_{\rho\pi}(s) = \sum_{V=\omega,\phi,\omega^\prime,{\ldots} }
 {g_{\gamma V}g_{V\rho\pi} \over D_V(s)}e^{i\phi_{\omega V}},
\end{eqnarray}
\begin{eqnarray}
 A_{\omega\pi}(s) = \sum_{V=\rho,\rho^\prime,{\ldots} }
 {g_{\gamma V}g_{V\omega\pi^0} \over D_V(s)}e^{i\phi_{\rho V}},
\end{eqnarray}
 where
$$D_V(s)=m_V^2-s-i\mbox{~}\sqrt[]{s}\Gamma_V(s), \mbox{~~~}
  \Gamma_V(s)=\sum_{f}\Gamma(V\to f,s).$$
 Here $f$ denotes the final state of the vector meson $V$ decay.
 $\phi_{\omega V}$ ($\phi_{\rho V}$) are relative interference phases
 between vector mesons $V$ and $\omega$ ($\rho$), so 
 $\phi_{\omega\omega}=0$ and
 $\phi_{\rho\rho}=0$. The coupling constants are determined through the decay
 branching ratios in the following way:
\begin{eqnarray}
\label{g}
 |g_{V\gamma}| = \Biggl[ {{3m_V^3\Gamma_VB(V \to e^+e^-)} \over
 {4\pi\alpha}} \Biggr]^{1/2}
\end{eqnarray}
\begin{eqnarray}
 |g_{V\rho\pi}| = \Biggl[{{4\pi\Gamma_VB(V \to \rho\pi)} \over
 {W_{\rho\pi}(m_V)}} \Biggr]^{1/2}, \mbox{~~}
\end{eqnarray}
\begin{eqnarray}
  |g_{V\omega\pi}| = \Biggl[{{12\pi\Gamma_VB(V \to \omega\pi)} \over   
  {q_{\omega\pi}^3(m_V)}} \Biggr]^{1/2},
\end{eqnarray}
 where $q_{\omega\pi}(s)$ is the $\omega$-meson momentum.

\section{Cross section measurement}

 From the data in Table~\ref{tab1} the cross section of the process
 $e^+e^-\to\pi^+\pi^-\pi^0$ can be calculated as follows:
\begin{eqnarray}
\label{aprox}
 \sigma(s) = {{N_{3\pi}(s)} \over {IL(s)\xi(s)}},
\end{eqnarray}
 where $N_{3\pi}(s)$ is the number of selected
 $e^+e^-\to\pi^+\pi^-\pi^0(\gamma)$ events, $IL(s)$ is the integrated
 luminosity, $\xi(s)$ is the function which takes into account the detection
 efficiency and radiative corrections for initial state radiation:
\begin{eqnarray}
\label{xifu}
 \xi(s) = {\int\limits^{E^{max}_\gamma}_0 \sigma_{3\pi}(s,E_\gamma)F(s,E_\gamma)
 \epsilon(s,E_\gamma) \mathrm{d}E_\gamma \over {\sigma_{3\pi}(s)}}.
\end{eqnarray}
 Here $E_\gamma$ is the emitted photon energy, $F(s,E_\gamma)$ is the electron
 ``radiator'' function \cite{fadin}, $\epsilon(s,E_\gamma)$ is the detection 
 efficiency of the process $e^+e^-\to\pi^+\pi^-\pi^0(\gamma_{rad})$ as a
 function of the emitted photon energy and the energy in the $e^+e^-$ 
 center of mass system, $\sigma_{3\pi}(s)$ is the theoretical energy dependence
 of the cross section given by equation (\ref{sech3p}).

 To obtain the values of $\xi(s)$ at each energy point, the visible cross
 section of the process $e^+e^-\to\pi^+\pi^-\pi^0(\gamma_{rad})$
 $$\sigma^{vis}(s) = {N_{3\pi}(s) \over IL(s)}$$
 was fitted by theoretical energy dependence
 $$\sigma^{th}(s) = \sigma_{3\pi}(s)\xi(s).$$
 The following logarithmic likelihood function was minimized:
 $$\chi^2=\sum_{i} {{(\sigma^{vis}_i-\sigma^{th}_i)^2}\over{\sigma^2_i}},$$
 where $i$ is the energy point number, $\sigma_i$ is the error of the visible
 cross section $\sigma^{vis}$.

 In a good approximation the contributions $\sigma_{\omega\pi\to3\pi}$ and
 $\sigma_{int}$ in expression (\ref{sech3p}) can be omitted, as they are
 rather small ($\sim$ 5 -- 10 \%) and actually do not modify the shape of
 $\sigma_{3\pi}(s)$ energy dependence. So we assumed that 
 $\sigma_{3\pi}(s)=\sigma_{\rho\pi\to3\pi}(s)$. The amplitude of the
 $\gamma^\star \to \rho\pi$ transition (\ref{aropi}) was written as
\begin{eqnarray}
 A_{\rho\pi}(s) = {{1}\over\sqrt[]{4\pi\alpha}}
 \sum_{V=\omega,\phi,\omega^\prime,\omega^{\prime\prime}}
 {{\Gamma_V m_V^2 \mbox{~} \sqrt[]{m_V\sigma(V\to 3\pi)}}\over{D_V(s)}}
 {{e^{i\phi_{\omega V}}}\over{\sqrt[]{W_{\rho\pi}(m_V)}}},
\end{eqnarray}
 where
$$ \sigma(V\to X) = {{12\pi B(V\to e^+e^-)B(V\to X) } \over {m_V^2}}.$$
 The following form of the energy dependence of the $\omega^\prime$ and
 $\omega^{\prime\prime}$ total width was used
 $$\Gamma_V(s)=\Gamma_V{W_{\rho\pi}(s)\over W_{\rho\pi}(m_V)}.$$
 In the fit the $\omega$ meson parameters (mass, width, branching ratios of
 main decays ) were fixed at their PDG values \cite{pdg}, and
 the $\phi$ meson mass and width were fixed at the values measured by SND
 \cite{phi98}. It was shown  \cite{phi98} that the
 $\sigma(\phi\to3\pi)$ parameter and the cross section value at
 $\sqrt[]{s} > 1027$ MeV
 have a rather large model error, due to the uncertainty in the choice of
 the phase $\phi_{\omega\phi}$ and the value of additional, besides the $\phi$
 and $\omega$ resonances, contributions to the transition amplitude. Therefore
 we have taken the $\sigma(\phi\to3\pi)$ as a free
 parameter in the fit and the visible cross section presented in this work
 was fitted together with the visible cross section  from Ref.\cite{phi98}.
 The masses and width of the $\omega^\prime$, $\omega^{\prime\prime}$
 resonances were free parameters of the fit. Phases $\phi_{\omega V}$ can
 deviate from $180^\circ$ or $0^\circ$ and their values can have energy
 dependence due to mixing between vector mesons. For example, the phase
 $\phi_{\omega\phi}$ was found to be close to $180^\circ$ \cite{phi98} and
 agree with the prediction \cite{faza}
 $\phi_{\omega\phi}=\Phi(s)$ $(\Phi(m_\phi)\simeq 163^\circ)$,
 where the function $\Phi(s)$ is defined in  Ref.\cite{faza}. There are no
 theoretical predictions of  $\phi_{\omega\omega^\prime}$ and
 $\phi_{\omega\omega^{\prime\prime}}$ values and their energy dependences, 
 and we have considered
 $\sqrt[]{\sigma(\omega^\prime\to 3\pi)}$ and
 $\sqrt[]{\sigma(\omega^{\prime\prime}\to3\pi)}$ as free parameters, i.e.
 $\phi_{\omega\omega^\prime}$ and $\phi_{\omega\omega^{\prime\prime}}$ can
 be equal to 0 or 180 degrees. The $\xi(s)$ values were obtained by
 approximation of the experimental data  in several models:
\begin{enumerate}
\item
 $\phi_{\omega\phi}=180^\circ$
\item
 $\phi_{\omega\phi}=\Phi(s)$
\item
 $\phi_{\omega\phi}$ is a free parameter
\item
 $\sigma(\omega^{\prime\prime}\to3\pi)=0$,
 $\phi_{\omega\phi}=180^\circ$
\item
 $\sigma(\omega^{\prime\prime}\to3\pi)=0$,
 $\phi_{\omega\phi}=\Phi(s)$
\item
 $\sigma(\omega^{\prime\prime}\to3\pi)=0$,
 $\phi_{\omega\phi}$ is a free parameter.
\end{enumerate}
 The values of $\xi(s)$ significantly depend on the applied model in
 the energy range
 $\sqrt[]{s} \simeq$ 1040 -- 1090 MeV, and at $\sqrt[]{s} = 1040$ MeV
 the $\xi(s)$ values differ by a factor 10 for different models.
 Above 1090 MeV the $\xi(s)$ model dependence is negligible.
 Using obtained $\xi(s)$ values, the cross section of the
 $e^+e^-\to\pi^+\pi^-\pi^0$ process was calculated (Table.\ref{tab2} ).
 The cross section in the energy region $\sqrt[]{s}=1027$ -- $1060$ MeV
 has changed in comparison with the values reported in Ref.\cite{phi98}.
 In Ref.\cite{phi98} contributions from the $\omega$ excitations were taken
 into account as a constant amplitude. In present analysis the more realistic
 model was used and it caused a change in the cross section.
 The systematic error of the cross section determination at each energy point
 $\sqrt[]{s}$ is equal to
 $$ 
 \sigma_{sys} = \sigma_{eff} \oplus \sigma_{IL} \oplus \sigma_{mod}(s) \oplus
 \sigma_{bkg}(s).
 $$
 Here $\sigma_{eff}=5\%$ and $\sigma_{IL}=2\%$ are systematic uncertainties
 in the detection efficiency and integrated luminosity, which are common for
 all energy points. The model uncertainty $\sigma_{mod}(s)$ is significant in 
 the region
 $\sqrt[]{s} =1027$ -- $1080$ MeV and was obtained from the difference of
 $\xi(s)$ values determined for the six models mentioned above. The error
 $\sigma_{bkg}(s)$ takes into account the inaccuracy ($\sim 40\%$) of
 background subtraction and depends on the beam energy. 

 The obtained cross section differs by about  $30\pm15$\% from the previous SND
 result \cite{sndmhad} (Fig.\ref{sis}), which claimed a systematic error 
 about 12\%.
 This difference is attributed to the fact that in the new  analysis we
 implemented corrections to the detection efficiency (described in III.C)
 which were not used in the previous
 one. The comparison of the measured cross section with the other experimental
 results is presented in Fig.\ref{cs}.

\section{Approximation of the $\pi^+\pi^-$ mass spectra.}
 The contribution of the $e^+e^-\to\omega\pi^0\to\rho^0\pi^0\to\pi^+\pi^-\pi^0$
 mechanism to the process $e^+e^-\to\pi^+\pi^-\pi^0$ is seen as the
 interference in the $\pi^+\pi^-$ invariant mass spectra. To analyze the
 dipion mass spectra, the formfactor $F$ (expression (\ref{formfac})) was
 presented in the following form:
\begin{eqnarray}
\label{formfac2}
 |F|^2 = \Biggl| A_{\rho\pi}(s) \Biggr|^2 \times \Biggl|  \sum_{i=+,0,-} 
 { g_{\rho^i\pi\pi} \over D_\rho(m_i)} + R(s)e^{i\psi(s)}
 {\mbox{Re}(\Pi_{\rho\omega}) g_{\rho^0\pi\pi} \over
 D_\rho(m_0) D_\omega(m_0)} \Biggr|^2,
\end{eqnarray}
 where $R(s)$ is the absolute value, and $\psi(s)$ is the phase of the
 ratio $A_{\omega\pi}(s)/A_{\rho\pi}(s)$. The $\psi(s)$ energy
 dependence can be obtained from the approximation of the experimental 
 $\pi^+\pi^-$
 invariant mass spectra as described below. The $R(s)$ value was calculated 
 from the equation
\begin{eqnarray}
 R^2\cdot \biggl(W_{\omega\pi}(s)-{q^3_{\omega\pi}(s)\over 3}
 {\sigma_{3\pi}(s)\over \sigma_{\omega\pi}(s)}\biggr) + 
 R\cdot \biggl(e^{-i\psi}W_{int}(s)+e^{i\psi}W_{int}^\star (s) \biggr)+
 W_{\rho\pi}(s) = 0,
\end{eqnarray}
 which follows from expressions (\ref{sech1}-\ref{sech3}). The
 $e^+e^-\to\omega\pi^0$ cross section was obtained from SND measurements
 of the $e^+e^-\to\omega\pi^0\to\pi^0\pi^0\gamma$ cross section \cite{ppg}:
 $\sigma_{\omega\pi^0}=\sigma_{\omega\pi^0\to\pi^0\pi^0\gamma}/
 B(\omega\to\pi^0\gamma)$,
 $\sigma_{3\pi}(s)$ is the $e^+e^-\to\pi^+\pi^-\pi^0$ cross section measured 
 here (Table \ref{tab2}).
 
 The real part of the polarization operator $\Pi_{\rho\omega}$ is proportional
 to $\sqrt[]{B(\omega\to\pi^+\pi^-)}$. The world average value for this 
 branching ratio is 
 $B(\omega\to\pi^+\pi^-)=2.21 \pm 0.30 \%$ \cite{pdg}. The results of
 $B(\omega\to\pi^+\pi^-)$ measurements in different experiments deviate from
 each other by a factor of more than 1.5 . For example, OLYA detector reported
 the value $B(\omega\to\pi^+\pi^-)=2.3\pm0.5 \%$ \cite{olya}, while CMD-2
 experiment reported $B(\omega\to\pi^+\pi^-)=1.33\pm0.25 \%$ \cite{cmdpp}.
 So $B(\omega\to\pi^+\pi^-)$ was considered as a free parameter of the fit.

 For the mass spectra analysis the events selected in the energy region
 $\sqrt[]{s} \ge 1100$ MeV were used. For each energy point the $\pi^+\pi^-$
 mass spectra were formed and arranged in histograms with a dipion mass range
 from 280 to 1240 MeV and bin width of 40 MeV. The invariant mass values were
 calculated after the kinematic reconstruction. The expected background was
 subtracted bin by bin while forming the desired histograms. 

 The analysis of the dipion mass spectra was performed in a way similar to this
 described in Ref.\cite{dplphi98}. The experimental spectra were fitted with
 theoretical distributions. Using the $e^+e^-\to\pi^+\pi^-\pi^0$ cross section
 (\ref{ds}) and formfactor (\ref{formfac2}), the theoretical spectra were
 calculated:
\begin{eqnarray}
 S^{(0)}_j(s) = {1 \over C_S(s)} \cdot\int\limits^{m_{j}+\Delta}_{m_{j}-\Delta}
 m_0 dm_0 \int\limits^{m_+^{max}(m_0)}_{m_+^{min}(m_0)}
  m_+ dm_+ |\vec{p}_+ \times \vec{p}_-|^2 \cdot |F|^2,
\end{eqnarray}
 where $j$ is the bin number, $\Delta=20$ MeV is a half of the bin width,
 $m_j$ is the central value of the invariant mass in the $j$th bin, $C_S(s)$ is
 a normalizing coefficient. These spectra were corrected taking into account
 the detection efficiency $\epsilon^{(0)}_j$ for the $j$th bin and a
 probability $a^{(0)}_{ij}$ for the event belonging to the $j$th bin to
 migrate to the $i$th bin due to the finite detector resolution
\begin{eqnarray}
 G^{(0)}_i(s) = {1 \over C_G(s)} \Biggl(\sum_j a^{(0)}_{ij} S^{(0)}_j(s)
 \epsilon^{(0)}_j \Biggr) \cdot (1+\delta^{(0)}_i(s)).
\end{eqnarray}
 Here $\delta^{(0)}_i(s)$ is a radiative correction and $C_G(s)$ is a
 normalizing coefficient. The values of $a^{(0)}_{ij}$, $\epsilon^{(0)}_{j}$
 and $\delta^{(0)}_i(s)$ were obtained from simulation.

 The function to be minimized was
\begin{eqnarray}
 \chi^2=\sum_{s} \chi^2_0(s) = \sum_{s} \sum_{i} \Biggl(
 { {H_i^{(0)}-G_i^{(0)}} \over {\sigma_i^{(0)}} }\Biggr)^2.
\end{eqnarray}
 Here $H^{(0)}$ is the normalized experimental $\pi^+\pi^-$ mass distribution
 (histogram);
 $\sigma_i^{(0)}=\Delta H^{(0)}_i \oplus \Delta G^{(0)}_i$  include the
 uncertainties $\Delta H^{(0)}_i$ and $\Delta G^{(0)}_i$ of the experimental
 and theoretical distributions  ($\Delta H^{(0)}_i \gg \Delta G^{(0)}_i$).

 During the fitting the phase $\psi(s)$ at each energy point and
 $B(\omega\to\pi^+\pi^-)$ were free parameters. Values of the phase 
 $\psi(s)$ 
 were allowed to vary from $-180^\circ$ to $180^\circ$. The obtained $\psi(s)$
 values are presented in Table~\ref{tab3}. The systematic inaccuracy of
 $\psi(s)$ is about $7^\circ$ and is connected with a systematic error in
 $R(s)$ determination, which in its turn is about $4\%$ due to uncertinities of
 $\sigma_{\omega\pi}$ and $\sigma_{3\pi}$ measurements. The
 $\omega\to\pi^+\pi^-$ decay probability was found to be equal to
 $2.38\pm^{1.77}_{0.90}\pm0.18 \%$,
 where the systematic error is also related to the uncertainty of the
 $R(s)$ determination.
 In Figs.\ref{neu} and \ref{cha} the experimental $\pi\pi$ mass spectra
 together with the theoretical distributions obtained from the fit and the 
 spectra expected from the only $\rho\pi$ intermediate state model are shown.
 In the $\pi^+\pi^-$ mass spectra
 the peak in the $\omega$ meson region is clearly seen. The distribution
 of the invariant mass
 of the $\pi^\pm\pi^0$ pairs does not contradict to  the $\rho\pi$ intermediate
 state model at the level of
 our statistical accuracy. These figures demonstrate that together with the
 $\rho\pi$ intermediate state the $\omega\pi^0$ intermediate state also
 contributes to the process $e^+e^-\to\pi^+\pi^-\pi^0$.

\section{The \lowercase{\boldmath{$e^+e^- \to\pi^+\pi^-\pi^0$}} total cross
         section analysis.}

 The analysis of the $e^+e^-\to \pi^+\pi^-\pi^0$ cross section energy
 dependence obtained here (Table~\ref{tab2}) met the following difficulties:
\begin{enumerate}
\item
 The cross section was measured in the limited $\sqrt[]{s}$ energy region
 and it is necessary to use the results of other experiments. As a result,
 because of different systematic effects the problem of matching cross sections
 of various measurements arises;
\item
 In the ideal case, to obtain the vector mesons parameters, the combined fit
 of all $e^+e^-\to hadrons$ cross sections is necessary;
\end{enumerate}

 The cross section measured in this work was analyzed together with the DM2
 results of the $e^+e^-\to \pi^+\pi^-\pi^0$ and $\omega\pi^+\pi^-$ \cite{dm2}
 cross sections measurements. The
 $e^+e^-\to \pi^+\pi^-\pi^0$ cross section was fitted by the expression
 (\ref{sech3p}). The $A_{\rho\pi}$ amplitude was written in the following way:
\begin{eqnarray}
 A_{\rho\pi}(s) = {{1}\over\sqrt[]{4\pi\alpha}}
 \Biggl(  { {\Gamma_\omega m_\omega^2 \mbox{~} 
 \sqrt[]{m_\omega\sigma(\omega\to 3\pi)}} \over {D_\omega(s)}}
 { 1 \over {\sqrt[]{W_{\rho\pi}(m_\omega)}}} +
 { {\Gamma_\phi m_\phi^2 \mbox{~} 
 \sqrt[]{m_\phi\sigma(\phi\to 3\pi)}} \over {D_\phi(s)}}
 { {e^{i\Phi(s)}} \over {\sqrt[]{W_{\rho\pi}(m_\phi)}}} +
\nonumber \\
 + \sum_{i=1}^3
 {{\Gamma_{\omega^i} m_{\omega^i}^2 \mbox{~} 
 \sqrt[]{m_{\omega^i}\sigma(\omega^i\to 3\pi)}}\over{D_{\omega^i}(s)}}
 {{e^{i\phi_{\omega\omega^i}}}\over{\sqrt[]{W_{\rho\pi}(m_{\omega^i})}}}
 \Biggr),
\end{eqnarray}
 where $i$ is the resonance number. The following form of the 
energy dependence of the $\omega^i$ total
 widths was used:
\begin{eqnarray}
 \Gamma_{\omega^1}(s)=\Gamma_{\omega^1}{W_{\rho\pi}(s)\over
 W_{\rho\pi}(m_{\omega^1})},
\end{eqnarray}
\begin{eqnarray}
 \Gamma_{\omega^i}(s)=\Gamma_{\omega^i}\biggl(B(\omega^i\to 3\pi)
 {W_{\rho\pi}(s) \over W_{\rho\pi}(m_{\omega^i})} + 
 B(\omega^i \to \omega\pi\pi) 
 {W_{\omega\pi\pi}(s) \over W_{\omega\pi\pi}(m_{\omega^i})}\biggr), 
 \mbox{~~~}
 i=2,3.
\end{eqnarray}
 Here $W_{\omega\pi\pi}(s)$ is the phase space factor of the $\omega\pi\pi$
 final state \cite{ak2}. The probabilities of the $\omega^i$ decays into
 $\pi^+\pi^-\pi^0$ and $\omega\pi\pi$  were calculated in the following way:
\begin{eqnarray}
 B(\omega^i\to f) = {\sigma(\omega^i\to f) \over 
\sum_f \sigma(\omega^i\to f)}. 
\end{eqnarray}
 Here $\sigma(\omega^i\to\omega\pi\pi) = 1.5 \cdot 
\sigma(\omega^i\to\omega\pi^+\pi^-)$.
 In the total width energy dependence the contributions from the following
 final states were neglected: $K_SK^\pm\pi^\mp$,
 $K^{\star0}K^-\pi^+$, $\overline{K}^{\star0}K^+\pi^-$, $K\overline{K}$.
 The $\omega$ meson parameters were fixed according to the PDG table values
 \cite{pdg}. The $m_\phi$, $\Gamma_{\phi}$ and parameters of the
 $\phi\to K\overline{K}$ and
 $\eta\gamma$ decays were fixed at the values obtained by SND \cite{phi98},
 while $\sigma(\phi\to 3\pi)$ was a free parameter of the fit. As it was
 mentioned above, the phases $\phi_{\omega\omega^i}$ can differ from 0 or 180
 degrees and be energy dependent. Here we consider only
 $\sqrt[]{\sigma(\omega^i\to 3\pi)}$ as a free parameter, i.e.
 $\phi_{\omega\omega^i}=0^\circ$ or $180^\circ$.

 For the $A_{\omega\pi}$ amplitude two models with different energy 
 behavior of the phase were used. Their parameters were obtained 
 by fitting the $e^+e^-\to\omega\pi^0\to\pi^0\pi^0\gamma$ cross section
 measured by SND \cite{ppg} and CLEO2 data on $\tau\to 3\pi\pi^0$ decay
 \cite{cleo4p} (Fig.\ref{omp}). The first model was suggested in 
 Ref. \cite{ppg}. It assumes that only the $\rho$ and
 $\rho^{\prime\prime}$ resonances contribute to the $e^+e^-\to\omega\pi$ cross
 section (i.e.
 $A_{\omega\pi}=A_{\rho\to\omega\pi}+ A_{\rho^{\prime\prime}\to\omega\pi}$),
 at that the following parameters are used: the coupling constant 
 $g_{\rho\omega\pi}\sim 15.2$ GeV$^{-1}$,$\rho^{\prime\prime}$-mass
 $m_{\rho^{\prime\prime}}\sim 1700$ MeV, width
 $\Gamma_{\rho^{\prime\prime}}\sim 1$ GeV, phase
 $\phi_{\rho\rho^{\prime\prime}}=180^\circ$ and
 $\sigma(\rho^{\prime\prime}\to\omega\pi) \sim 9$ nb. The $\rho^{\prime\prime}$
 total width energy dependence is taken to be the following
\begin{eqnarray}
 \Gamma_{\rho^{\prime\prime}}(s)=\Gamma_{\rho^{\prime\prime}} \biggl(
 0.1{m^2_{\rho^{\prime\prime}} \over s} {q^3_{\pi\pi}(s) \over
 q^3_{\pi\pi}(m_{\rho^{\prime\prime}})} + 
 0.9{ q^3_{\omega\pi}(s) \over q^3_{\omega\pi}(m_{\rho^{\prime\prime}})}
 \biggr), 
\end{eqnarray}
 where $q_{\pi\pi}(s)$ is the pion momentum. The second model assumes that
 three $\rho$, $\rho^\prime$ and $\rho^{\prime\prime}$ resonances
 contribute to the $e^+e^-\to\omega\pi$ cross section  (i.e.
 $A_{\omega\pi}=A_{\rho\to\omega\pi}+ A_{\rho^\prime\to\omega\pi}+
 A_{\rho^{\prime\prime}\to\omega\pi}$). In this case the parameters 
 of the model are:
 $g_{\rho\omega\pi}\sim 16.8$ GeV$^{-1}$,
 $m_{\rho^{\prime}}\sim 1480$ MeV, $\Gamma_{\rho^{\prime}}\sim 790$ MeV,
 $\phi_{\rho\rho^\prime}=180^\circ$,
 $\sigma(\rho^\prime\to\omega\pi) \sim 86$ nb, and
 $m_{\rho^{\prime\prime}}\sim 1640$ MeV,
 $\Gamma_{\rho^{\prime\prime}}\sim 1290$ MeV,
 $\phi_{\rho\rho^{\prime\prime}}=0^\circ$,
 $\sigma(\rho^{\prime\prime}\to\omega\pi) \sim 48$ nb. The $\rho^\prime$ and
 $\rho^{\prime\prime}$ total width energy dependence is taken in the form
\begin{eqnarray}
 \Gamma_{\rho^{\prime(\prime\prime)}}(s)=\Gamma_{\rho^{\prime(\prime\prime)}}
 { q^3_{\omega\pi}(s) \over q^3_{\omega\pi}(m_{\rho^{\prime(\prime\prime)}})}
\end{eqnarray}
 In both models the $\rho$ meson energy dependent width has the form:
\begin{eqnarray}
 \Gamma_{\rho}(s)=\Gamma_{\rho^0} {{m_{\rho^0}^2}\over{s}}
 {{q^3_{\pi\pi}(s)}\over{q^3_{\pi\pi}(m_{\rho^0})}} +
 {{g_{\rho\omega\pi}^2}\over{12\pi}}q^3_{\omega\pi}(s)
\end{eqnarray}

 The $e^+e^-\to\omega\pi^+\pi^-$ process cross section was written in the
 following way:
\begin{eqnarray}
 \sigma_{\omega\pi\pi} = {1 \over s^{3/2}} \Biggl| \sum_{i=2}^3 
 {{\Gamma_{\omega^i}m_{\omega^i}^2\mbox{~}\sqrt[]{\sigma(\omega^i\to\omega\pi^+\pi^-)
 m_{\omega^i}}} \over{D_{\omega^i}(s)}} \mbox{~}
 \sqrt[]{{W_{\omega\pi\pi}(s)}\over{W_{\omega\pi\pi}(m_{\omega^i})}} \Biggr|^2.
\end{eqnarray}

 The cross sections of the $e^+e^-\to \pi^+\pi^-\pi^0$ and
 $\omega\pi^+\pi^-$  processes measured by SND and DM2 were fitted together.
 The function to be minimized was
 $$\chi^2=\chi^2_{3\pi(SND)}+\chi^2_{3\pi(DM2)}+\chi^2_{\omega\pi\pi(DM2)},$$
 where
 $$\chi^2_{3\pi(SND)} = \sum_{s}\Biggl( 
 { {\sigma_{3\pi}^{(SND)}(s)-\sigma_{3\pi}(s)} \over
{\Delta_{3\pi}^{(SND)}(s)} }
 \Biggr)^2$$
 $$\chi^2_{3\pi(DM2)} = \sum_{s}\Biggl( 
 { C_{3\pi}\cdot{\sigma_{3\pi}^{(DM2)}(s)-\sigma_{3\pi}(s)} \over
{\Delta_{3\pi}^{(DM2)}(s)} }
 \Biggr)^2$$
 $$\chi^2_{\omega\pi\pi(DM2)} = \sum_{s}\Biggl(
 { C_{\omega\pi\pi}\cdot{\sigma_{\omega\pi\pi}^{(DM2)}(s)-
 \sigma_{\omega\pi\pi}(s)} \over {\Delta_{\omega\pi\pi}^{(DM2)}(s)} }
 \Biggr)^2$$
 Here $\sigma_{3\pi(\omega\pi\pi)}^{(SND(DM2))}(s)$ are the experimental cross
 sections, $\Delta$ are their uncertainties, $C_{3\pi}$ and $C_{\omega\pi\pi}$
 are coefficients which take into account the relative systematic bias
 between SND and DM2 data. The $e^+e^-\to\pi^+\pi^-\pi^0$ cross section
 measured by SND (Table~\ref{tab2}) was fitted in the energy region
 $\sqrt[]{s}$ from 980 to 1380 MeV. The errors $\Delta_{3\pi(SND)}$ include
 the statistical $\sigma_{stat}$ and the following systematic errors:
 $\sigma_{bkg}$ due to the
 inaccuracy of the background subtraction and $\sigma_{mod}$ due to model
 dependence. Thus
 $\Delta_{3\pi(SND)}=\sigma_{stat}\oplus\sigma_{mod}\oplus\sigma_{bkg}$.
 The fitting was performed with $m_{\omega^i}$, $\Gamma_{\omega^i}$,
 $\sqrt[]{\sigma(\omega^i\to 3\pi)}$,
 $\sqrt[]{\sigma(\omega^i\to\omega\pi^+\pi^-)}$ and $\sigma(\phi\to 3\pi)$
 as free parameters.

 To estimate the possible relative bias between SND and DM2 data, the
 $C_{3\pi}$ considered as a free parameter as well. It was found that
 $C_{3\pi}=1.72\pm 0.24$. To estimate the possible biases independently
 the cross sections of the $e^+e^-\to\omega\pi^0$ process (Fig.\ref{omp})
 measured by
 SND \cite{ppg} and DM2 \cite{dm2omp}, and cross section calculated, by using
 CVC hypothesis, from the CLEO2 result on the $\tau\to 3\pi\pi^0$ decay 
 \cite{cleo4p} were also studied. The $e^+e^-\to\omega\pi^0$ cross section 
 was measured by DM2
 by using $\pi^+\pi^-2\pi^0$ final state, i.e. as in the case of the 
 $\pi^+\pi^-\pi^0$ and $\omega\pi^+\pi^-$ final states the events containing
 both  tracks and photons were detected. This gives us a hope that all these 
 DM2 measurements have similar systematic errors. The SND and CLEO2 data agree
 rather well.  The DM2 and CLEO2 data points are strongly overlapped. 
 The average
 ratio of the CLEO2 and DM2 cross sections is 1.54, and this agrees with
 $C_{3\pi}=1.72\pm 0.24$. In further analysis we assumed
 $C_{3\pi}=C_{\omega\pi\pi}$ and fixed these coefficients at 1 or 1.54.

 It is generally accepted that two $\omega$-like resonances $\omega^\prime$
 and $\omega^{\prime\prime}$ exist \cite{pdg,dm2}.
 The first fit was done by assuming that the number of the $\omega^i$
 resonances is equal to 3 and without taking into account the 
 $\omega\pi\to\pi^+\pi^-\pi^0$ mechanism (i.e., $\sigma_{\omega\pi\to 3\pi}=0$
 and $\sigma_{int}=0$ were assumed). The obtained parameters of the $\omega^i$
 resonances are shown in Table~\ref{tab44}. The $\sigma(\omega^1\to 3\pi)$ differs from zero by
 about one standard deviation. If in this approximation one takes into account
 the contribution from the $\omega\pi\to\pi^+\pi^-\pi^0$ mechanism, then 
 $\sigma(\omega^1\to 3\pi)= 0.07 \pm^{0.32}_{0.07}$ nb, and the parameters
 of the $\omega^2$,
 $\omega^3$ resonances deviate from their previous values within their 
 statistical errors.
 So in the further analysis the parameter $\sigma(\omega^1\to 3\pi)$ was fixed
 to zero and for the $\omega^2$, $\omega^3$ resonances a more usual notation
 $\omega^\prime$, $\omega^{\prime\prime}$ was used.

 The further fittings were performed under the following assumptions:
\begin{enumerate}
\item
 the contribution from the $\omega\pi\to\pi^+\pi^-\pi^0$ was not taken into
 account, i.e. 
 $\sigma_{\omega\pi\to 3\pi}=0$, $\sigma_{int}=0$.
\item
 the first model for the amplitude $A_{\omega\pi}$ was used;
\item
 the second model for the amplitude $A_{\omega\pi}$ was used;
\end{enumerate}
 The results of the fits are shown in Tables~\ref{tab4}, \ref{tab5} and
 Fig.\ref{cs3}, \ref{cs4}. In case when no relative shift between SND and
 DM2 experiments was assumed, the value of $\chi^2_{3\pi(DM2)}$ is rather
 large. The obtained parameters depend weakly on the applied model.

\section{Discussion}
 The fit results revealed that the $e^+e^-\to\pi^+\pi^-\pi^0$ and
 $e^+e^-\to\omega\pi^+\pi^-$ cross sections can be described by a sum of
 contributions of the 
 $\omega$ and $\phi$ mesons and two additional $\omega^\prime$, 
 $\omega^{\prime\prime}$
 resonances. The following  $\omega^\prime$ parameters were
 obtained (Table~\ref{tab4}):
$$
 m_{\omega^\prime} = 1490 \pm 50 \pm 25 \mbox{~~MeV},
$$
$$
 \Gamma_{\omega^\prime} = 1210 \pm^{300}_{200} \pm 170 \mbox{~~MeV},
$$
$$
 \sigma(\omega^\prime\to 3\pi) = 3.5 \pm 0.5 \pm 0.2 \mbox{~~nb},
$$
$$
 \sigma(\omega^\prime\to\omega\pi^+\pi^-) = 0.03 \pm^{0.1}_{0.03} \pm 0.01
\mbox{~~nb},
$$
$$
 \phi_{\omega\omega^\prime} \sim 180^\circ
$$ 
 The $\omega^{\prime}$ decays mostly into $\pi^+\pi^-\pi^0$:
 $B(\omega^2\to 3\pi) \simeq 99 \%$ and its electronic width is
 $\Gamma(\omega^\prime\to e^+e^-) \simeq 650$ eV. The $\omega^{\prime\prime}$
 parameters were found to be:
$$
 m_{\omega^{\prime\prime}} = 1790 \pm 40 \pm 10 \mbox{~~MeV},
$$
$$
 \Gamma_{\omega^{\prime\prime}} = 560 \pm^{150}_{100} \pm 20 \mbox{~~MeV},
$$
$$
 \sigma(\omega^{\prime\prime}\to 3\pi) = 2.0 \pm 0.40 \pm 0.8 
 \mbox{~~nb},
$$
$$
 \sigma(\omega^{\prime\prime}\to\omega\pi^+\pi^-) = 1.9 \pm 0.4 \pm 0.8
 \mbox{~~nb},
$$
$$
 \phi_{\omega\omega^{\prime\prime}} \sim 0^\circ
$$ 
 The $\omega^{\prime\prime}$ resonance decays with approximately equal
 probabilities into $\pi^+\pi^-\pi^0$ and $\omega\pi\pi$:
 $B(\omega^{\prime\prime}\to 3\pi)\simeq 0.4$,
 $B(\omega^{\prime\prime}\to \omega\pi\pi) \simeq 0.6$ and it has the 
 electronic width
 $\Gamma(\omega^{\prime\prime}\to e^+e^-) \simeq 600$ eV.
 The second errors shown are due to the uncertainty of the
 $A_{\omega\pi}$ amplitude choice and possible relative bias between different 
 experiments. 

 The rather large electronic widths obtained for the
 $\omega^{\prime}$ and $\omega^{\prime\prime}$ resonances may represent some
 challenge for theory. In the framework of the nonrelativistic quark
 model one can obtain the following  ratios:
 $$
 \biggl| {{\Psi_{\omega^{\prime}}^S(0)}\over{\Psi_\omega^S(0)}}\biggr|^2 =
 \biggl({{m_{\omega^\prime}}\over{m_\omega}} \biggr)^2 \cdot
 {{\Gamma(\omega^\prime\to e^+e^-)}\over{\Gamma(\omega\to e^+e^-)}}
 \sim 4,
 $$
 $$
 \biggl| {{\Psi_{\omega^{\prime\prime}}^S(0)}\over{\Psi_\omega^S(0)}}\biggr|^2
 = \biggl({{m_{\omega^{\prime\prime}}}\over{m_\omega}} \biggr)^2 \cdot
 {{\Gamma(\omega^{\prime\prime}\to e^+e^-)}\over{\Gamma(\omega\to e^+e^-)}}
 \sim 5,
 $$
 where $\Psi_V^S(0)$ is the radial wave function of the $q\overline{q}$ bound
 state at the origin. For the quark-antiquark potentials
 used to describe heavy quarkonia, such ratios are always less than
 unity \cite{zk1}. This is also confirmed experimentally. For example,
 analogous ratios for $c\overline{c}$ and $b\overline{b}$ states are:
 $|\Psi_{\psi(2S)}^S(0)/\Psi_{J/\psi}^S(0)|^2\simeq 0.57$,
 $|\Psi_{\Upsilon(2S)}^S(0)/\Psi_{\Upsilon(1S)}^S(0)|^2\simeq 0.44$,
 $|\Psi_{\Upsilon(3S)}^S(0)/\Psi_{\Upsilon(1S)}^S(0)|^2\simeq 0.43$. Of course,
 the nonrelativistic quark model is unreliable for light-quark $\omega$-states.
 But, surprisingly, it gives quite reasonable description of the ground state
 $\rho$, $\omega$ and $\phi$ meson leptonic widths, which do not change
 radically in the framework of the ``relativized'' quark model \cite{zk2}. For
 comparison, the nonrelativistic quark model predictions for the two photon
 widths of the light pseudoscalar mesons are dramatically wrong and only the
 ``relativized'' model gives reasonable result \cite{zk3}. More precise data
 and  more deep analysis is required to draw strict conclusions.

 The $\omega^\prime$, $\omega^{\prime\prime}$ widths obtained from the fit
 are rather large in comparison with their masses (this result agrees with
 experimental data analysis reported in \cite{ak1,ak2,ak3}). In this
 context the question whether the sum of Breit-Wigner amplitudes is an 
 adequate description of the cross sections in the energy region
 $m_\phi<\sqrt[]{s}<2000$ MeV becomes actual. 

 The presented analysis of the $\omega$-like excited states is somewhat
 speculative since we had to assume a rather large systematic bias between SND
 and DM2 measurements.

 The $\sigma(\phi\to 3\pi)$ was found to be equal
 $$\sigma(\phi\to 3\pi) = 646 \pm 4 \pm 37 \mbox{~nb}.$$
 This agrees with the results of SND studies of the $e^+e^-\to\pi^+\pi^-\pi^0$
 cross section in the vicinity of the $\phi$ resonance
 $\sigma(\phi\to 3\pi)=659\pm35$ nb \cite{phi98}. The slight deviations in the
 central value and the error can be related to the difference in descriptions 
 of the $\omega^\prime$, $\omega^{\prime\prime}$ contributions  used 
 in these works. The fit was performed by assuming
 $\phi_{\omega\phi}=\Phi(s)$
 \cite{faza}. If $\phi_{\omega\phi}$ is considered to be a free parameter of 
 the fit, then its value is:
 $$\phi_{\omega\phi}=164^\circ \pm 3^\circ,$$ 
 which agrees with $\Phi(m_\phi)=163^\circ$ \cite{faza}.

 The relative phase $\psi(s)$ between $A_{\rho\pi}$ and $A_{\omega\pi}$
 amplitudes and $B(\omega\to\pi^+\pi^-)$ was obtained from the $\pi^+\pi^-$
 invariant mass spectra analysis in the $\sqrt[]{s}$ energy region from 1100 to
 1380 MeV (Table~\ref{tab3}, Fig.\ref{faz2}). The phase $\psi(s)$ can be also 
 calculated from the total cross section fit results (Table~\ref{tab4}).
 Figure~\ref{faz2} demonstrates that the phase $\psi(s)$ energy dependence
 cannot be described if the model with $A_{\omega\pi}=A_{\rho\to\omega\pi}+
 A_{\rho^{\prime\prime}\to\omega\pi}$ is used. On the other hand, the 
 model in which $A_{\omega\pi}=A_{\rho\to\omega\pi}+
 A_{\rho^\prime\to\omega\pi}+A_{\rho^{\prime\prime}\to\omega\pi}$ gives 
 satisfactory description of the data. The $\omega\to\pi^+\pi^-$ decay
 probability was found to be:
 $$B(\omega\to\pi^+\pi^-)=2.38\pm^{1.77}_{0.90}\pm0.18 \%$$
 This result does not contradict both to OLYA measurements \cite{olya} and 
 world average
 value \cite{pdg}, as well as to  CMD2 result \cite{cmdpp}.
 Using the results of the total $e^+e^-\to\pi^+\pi^-\pi^0$ cross section and
 $\pi^+\pi^-$ invariant mass spectra analysis, the contribution of the 
 $e^+e^-\to\rho\pi\to \pi^+\pi^-\pi^0$ mechanism to the total cross section
 was estimated to be $\sim 90\%$
 in the energy range $\sqrt[]{s}=1100$ -- 1380 MeV.

 For the data analysis the model which takes into account only
 $e^+e^-\to\rho\pi\to\pi^+\pi^-\pi^0$ and $\omega\pi^0\to\pi^+\pi^-\pi^0$
 mechanisms were used. The 
 $e^+e^-\to\rho^{\prime(\prime\prime)}\pi\to\pi^+\pi^-\pi^0$, 
 intermediate state, as well as the $\rho$ and $\pi$
 meson interaction in the final state \cite{akfaz} are also possible.
 Taking into account these contributions in the fit can change the
 $\psi(s)$ values, but the statistics collected in SND experiments is not 
 enough for
 studies of such contributions. In addition, the parameters of the
 $\rho^{\prime(\prime\prime)}$ resonances are poorly established. In the energy
 dependence of the total width the contributions from the following decays were
 not taken into account: $\omega^{\prime(\prime\prime)}\to K_SK^\pm\pi^\mp,
 K^{\star0}K^-\pi^+(\overline{K}^{\star0}K^+\pi^-), K\overline{K}$,
 $\rho^{\prime(\prime\prime)}\to\rho\pi\pi, \eta\pi^+\pi^-, K\overline{K},
 K_SK^\pm\pi^\mp, K^{\star0}K^-\pi^+(\overline{K}^{\star0}K^+\pi^-)$.
 The mixing between vector mesons excitations was neglected. It is possible
 that a more detailed model for the $A_{\omega\pi}$ and $A_{\rho\pi}$
 amplitudes can change the  calculated energy dependence of the phase
 $\psi(s)$ presented in  Fig.\ref{faz2}.

 At present in BINP (Novosibirsk) the VEPP-2000 collider with energy range from
 0.36 to 2 GeV and luminosity up to $10^{32}$ cm$^{-2}$s$^{-1}$ (at$\sqrt[]{s}
 \sim 2$ GeV) is under construction \cite{vepp2000}. The two detectors SND
 \cite{sndupgrade} and CMD-2M \cite{cmd2m} are being upgraded for experiments 
 at this new facility. In these experiments the increase of the accuracy in
 determination of $e^+e^- \to hadrons$ cross sections is expected 
 in the energy range $m_\phi<\sqrt[]{s}<2000$
 MeV. We hope that the new data will improve the understanding of the nature
 of $\rho^{\prime(\prime\prime)}$, $\omega^{\prime(\prime\prime)}$ and
 $\phi^{\prime(\prime\prime)}$ resonances, as well as their decay mechanisms 
 and theoretical methods of their description.
 
\section{Conclusion}
 The cross section of the process $e^+e^-\to \pi^+\pi^-\pi^0$ was measured in
 the SND experiment at the VEPP-2M collider in the energy region
 $\sqrt[]{s} = 980$ -- 1380 MeV. Due to the increased luminosity, and improved
 corrections for analysis losses and initial state radiation, the cross section
 measurements reported here (Table~\ref{tab2}) supersede those in
 Ref.\cite{sndmhad} and Ref.\cite{phi98}.
 The measured cross section was analyzed in
 the framework of the generalized vector meson dominance model together with
 the $e^+e^-\to \pi^+\pi^-\pi^0$ and $\omega\pi^+\pi^-$ cross sections
 obtained by DM2. It was found that the experimental data can be described 
 with a sum of contributions of $\omega$, $\phi$ mesons and two
 $\omega^\prime$ and $\omega^{\prime\prime}$ resonances  with
 masses $m_{\omega^\prime}\sim 1490$, $m_{\omega^{\prime\prime}}\sim 1790$ MeV
 and widths $\Gamma_{\omega^\prime}\sim 1210$,
 $\Gamma_{\omega^{\prime\prime}}\sim 560$ MeV. The analysis of the
 dipion mass spectra in the energy region $\sqrt[]{s}$ from 1100 to 1380
 MeV has shown that for their description the mechanism 
 $e^+e^-\to\omega\pi^0\to\pi^+\pi^-\pi^0$ is required. The
 phase between $e^+e^-\to\omega\pi$ and $e^+e^-\to\rho\pi$ processes amplitudes
 was measured for the first time. Its value is close to zero and depends on
 energy.

\section*{acknowledgments}

 The authors are grateful to N.N.Achasov, S.I.Eidelman and A.A.Kozhevnikov for
 useful discussions. The present work was supported in part by grant no. 78
 1999 of Russian Academy of Science for young scientists and grant STP
 ``Integration'' A0100.

\newpage

\begin{table}
\caption{Event numbers $N_{3\pi}$ of the
         $e^+e^-\to\pi^+\pi^-\pi^0(\gamma)$ process (after background
	 subtraction) and $N_{bkg}$ of background processes, integrated
	 luminosity $IL$ and detection efficiency $\epsilon(s,E_\gamma=0)$
         (without $\gamma$-quantum radiation). $\delta_{rad}$ is radiative
	 correction ($\delta_{rad}=\xi(s)/\epsilon(s,E_\gamma=0)$, 
	 $\xi(s)$ is defined through the expression (\ref{xifu})).}
\label{tab1}
\begin{center}
\begin{tabular}[t]{cccccc}
$\sqrt[]{s}$ (MeV)&$IL$ (nb$^{-1}$)&$\epsilon(s,E_\gamma=0)$&
$\delta_{rad}$&$N_{3\pi}$&
$N_{bkg}$ \\ \hline
980&129&0.150&  0.858            &259$\pm$18& 3$\pm$1 \\
1040& 69&0.153&11.706 -- 131.646 & 90$\pm$10& 4$\pm$1 \\
1050& 84&0.149& 3.762 -- 5.281   & 75$\pm$10& 4$\pm$1 \\
1060&279&0.150& 1.808 -- 2.018   &196$\pm$16& 8$\pm$2 \\
1070& 98&0.150& 1.269 -- 1.327   & 61$\pm$ 9& 2$\pm$1 \\
1080&578&0.150& 1.060 -- 1.102   &325$\pm$23&22$\pm$6 \\
1090& 95&0.150& 0.985 -- 1.002   & 54$\pm$ 8& 3$\pm$1 \\
1100&445&0.152& 0.928            &255$\pm$18&14$\pm$3 \\
1110& 90&0.151& 0.915            & 70$\pm$11& 2$\pm$1 \\
1120&306&0.150& 0.889            &213$\pm$17&11$\pm$3 \\
1130&113&0.151& 0.889            & 76$\pm$10& 4$\pm$1 \\
1140&289&0.151& 0.901            &177$\pm$16& 9$\pm$2 \\
1150& 69&0.152& 0.873            & 59$\pm$ 9& 2$\pm$1 \\
1160&320&0.152& 0.877            &217$\pm$17&11$\pm$2 \\
1180&423&0.152& 0.884            &302$\pm$21&12$\pm$3 \\
1190&172&0.152& 0.872            &125$\pm$12& 4$\pm$1 \\
1200&439&0.153& 0.883            &290$\pm$19&13$\pm$2 \\
1210&151&0.153& 0.871            &129$\pm$12& 4$\pm$1 \\
1220&343&0.153& 0.947            &282$\pm$19& 9$\pm$2 \\
1230&141&0.153& 0.871            &103$\pm$11& 4$\pm$1 \\
1240&378&0.153& 0.871            &250$\pm$17& 6$\pm$1 \\
1250&209&0.154& 0.871            &165$\pm$14& 6$\pm$1 \\
1260&163&0.154& 0.867            &129$\pm$13& 5$\pm$1 \\
1270&241&0.154& 0.868            &175$\pm$15& 8$\pm$2 \\
1280&229&0.154& 0.872            &169$\pm$13& 8$\pm$2 \\
1290&272&0.155& 0.866            &199$\pm$15& 9$\pm$2 \\
1300&272&0.155& 0.867            &188$\pm$14& 6$\pm$2 \\
1310&202&0.155& 0.874            &153$\pm$14& 5$\pm$1 \\
1320&236&0.155& 0.873            &174$\pm$14& 7$\pm$2 \\
1330&293&0.156& 0.876            &206$\pm$15& 8$\pm$2 \\
1340&439&0.156& 0.874            &281$\pm$20&12$\pm$2 \\
1350&257&0.156& 0.876            &169$\pm$14& 6$\pm$2 \\
1360&625&0.156& 0.872            &399$\pm$22&19$\pm$3 \\
1370&256&0.156& 0.879            &179$\pm$15& 7$\pm$2 \\
1380&480&0.157& 0.880            &278$\pm$18&16$\pm$4 \\ \hline
\end{tabular}
\end{center}
\end{table}

\begin{table}
\caption{The $e^+e^-\to\pi^+\pi^-\pi^0$ cross section. $\star$ denotes
the points in which the cross section was calculated using data from
Ref. \cite{phi98} (the cross section has changed only for energies
$\sqrt[]{s}>1027$ MeV). $\sigma_{mod}$ is model uncertainty, $\sigma_{bkg}$ is
 the error due to background subtraction, $\sigma_{eff}\oplus\sigma_{IL}$ -
 error due to uncertainty in detection efficiency and integrated luminosity
 determination (5\%  at the energies marked by $\star$ and 5.4\% for other 
 energy points),
 $\sigma_{sys}=\sigma_{eff} \oplus \sigma_{IL} \oplus \sigma_{mod}(s) 
 \oplus \sigma_{bkg}(s)$
 is the total systematic error. }
\label{tab2}
\begin{center}
\footnotesize
\begin{tabular}[t]{cccccc}
$\sqrt[]{s}$(MeV)&$\sigma$(nb)&$\sigma_{mod}$(nb)&$\sigma_{bkg}$(nb)&
$\sigma_{eff}\oplus\sigma_{IL}$(nb)&$\sigma_{sys}$(nb)\\ \hline
 980.00        &15.58 $\pm$ 1.07& 0.00 &  0.00 &  0.84 &  0.84 \\
 984.02$^\star$&17.30 $\pm$ 0.80& 0.00 &  0.00 &  0.86 &  0.86  \\
 984.21$^\star$&18.10 $\pm$ 0.90& 0.00 &  0.00 &  0.91 &  0.91   \\
1003.71$^\star$&37.60 $\pm$ 1.40& 0.00 &  0.00 &  1.88 &  1.88    \\
1003.91$^\star$&36.20 $\pm$ 1.30& 0.00 &  0.00 &  1.81 &  1.81    \\
1010.17$^\star$&68.50 $\pm$ 2.40& 0.00 &  0.00 &  3.42 &  3.42    \\
1010.34$^\star$&69.50 $\pm$ 2.50& 0.00 &  0.00 &  3.48 &  3.48    \\
1015.43$^\star$&220.00$\pm$ 6.50& 0.00 &  0.00 & 11.00 & 11.00    \\
1015.75$^\star$&243.10$\pm$ 7.50& 0.00 &  0.00 & 12.16 & 12.16    \\
1016.68$^\star$&358.90$\pm$10.60& 0.00 &  0.00 & 17.94 & 17.94    \\
1016.78$^\star$&353.60$\pm$11.10& 0.00 &  0.00 & 17.68 & 17.68    \\
1017.59$^\star$&493.60$\pm$14.90& 0.00 &  0.00 & 24.68 & 24.68    \\
1017.72$^\star$&515.00$\pm$15.30& 0.00 &  0.00 & 25.75 & 25.75    \\
1018.62$^\star$&664.20$\pm$13.10& 0.00 &  0.00 & 33.21 & 33.21    \\
1018.78$^\star$&658.60$\pm$11.60& 0.00 &  0.00 & 32.93 & 32.93   \\ 
1019.51$^\star$&667.00$\pm$11.80& 0.00 &  0.00 & 33.35 & 33.35  \\  
1019.79$^\star$&595.50$\pm$14.10& 0.00 &  0.00 & 29.77 & 29.77  \\  
1020.43$^\star$&471.20$\pm$15.50& 0.00 &  0.00 & 23.56 & 23.56 \\
1020.65$^\star$&399.80$\pm$14.50& 0.00 &  0.00 & 19.99 & 19.99 \\
1021.41$^\star$&270.10$\pm$ 9.90& 0.00 &  0.00 & 13.51 & 13.51 \\   
1021.68$^\star$&217.40$\pm$ 8.50& 0.00 &  0.00 & 10.87 & 10.87 \\
1022.32$^\star$&142.90$\pm$ 6.10& 0.00 &  0.00 &  7.14 &  7.14 \\
1023.27$^\star$&92.20 $\pm$ 3.40& 0.00 &  0.00 &  4.61 &  4.61 \\
1027.52$^\star$&15.33 $\pm$ 0.73& 0.57 &  0.00 &  0.77 &  0.96 \\
1028.23$^\star$&10.81 $\pm$ 0.62& 0.52 &  0.00 &  0.54 &  0.75 \\
1033.58$^\star$&1.75  $\pm$ 0.11& 0.47 &  0.00 &  0.09 &  0.48 \\
1033.84$^\star$&1.43  $\pm$ 0.12& 0.41 &  0.00 &  0.07 &  0.42 \\
1039.59$^\star$&0.37  $\pm$ 0.04& 0.31 &  0.00 &  0.02 &  0.31 \\
1039.64$^\star$&0.37  $\pm$ 0.03& 0.31 &  0.00 &  0.02 &  0.31 \\
1040.00        &0.40  $\pm$ 0.04& 0.33 &  0.00 &  0.02 &  0.33 \\
1049.60$^\star$&1.12  $\pm$ 0.12& 0.20 &  0.00 &  0.06 &  0.21 \\
1049.81$^\star$&1.14  $\pm$ 0.15& 0.20 &  0.00 &  0.06 &  0.21 \\
1050.00        &1.37  $\pm$ 0.18& 0.23 &  0.00 &  0.07 &  0.24 \\
1059.52$^\star$&1.75  $\pm$ 0.21& 0.09 &  0.00 &  0.09 &  0.13 \\
1059.66$^\star$&1.84  $\pm$ 0.28& 0.09 &  0.00 &  0.09 &  0.13 \\
1060.00        &2.46  $\pm$ 0.20& 0.14 &  0.00 &  0.13 &  0.19 \\
1070.00        &3.21  $\pm$ 0.47& 0.07 &  0.04 &  0.17 &  0.19 \\
1080.00        &3.46  $\pm$ 0.24& 0.07 &  0.09 &  0.19 &  0.22 \\
1090.00        &3.84  $\pm$ 0.57& 0.03 &  0.07 &  0.21 &  0.22 \\
1100.00        &4.07  $\pm$ 0.29& 0.00 &  0.09 &  0.22 &  0.24 \\
1110.00        &5.66  $\pm$ 0.89& 0.00 &  0.05 &  0.31 &  0.31 \\
1120.00        &5.19  $\pm$ 0.42& 0.00 &  0.11 &  0.28 &  0.30 \\
1130.00        &5.04  $\pm$ 0.67& 0.00 &  0.10 &  0.27 &  0.29 \\
1140.00        &4.50  $\pm$ 0.40& 0.00 &  0.09 &  0.24 &  0.26 \\
1150.00        &6.40  $\pm$ 0.98& 0.00 &  0.10 &  0.35 &  0.36 \\
1160.00        &5.12  $\pm$ 0.39& 0.00 &  0.10 &  0.28 &  0.29 \\
1180.00        &5.30  $\pm$ 0.37& 0.00 &  0.09 &  0.29 &  0.30 \\
1190.00        &5.44  $\pm$ 0.53& 0.00 &  0.08 &  0.29 &  0.30 \\
1200.00        &4.89  $\pm$ 0.32& 0.00 &  0.09 &  0.26 &  0.28 \\
1210.00        &6.39  $\pm$ 0.60& 0.00 &  0.08 &  0.34 &  0.35 \\
1220.00        &5.68  $\pm$ 0.41& 0.00 &  0.07 &  0.31 &  0.32 \\
1230.00        &5.48  $\pm$ 0.59& 0.00 &  0.09 &  0.30 &  0.31 \\
1240.00        &4.96  $\pm$ 0.34& 0.00 &  0.04 &  0.27 &  0.27 \\
1250.00        &5.91  $\pm$ 0.51& 0.00 &  0.08 &  0.32 &  0.33 \\
1260.00        &5.92  $\pm$ 0.60& 0.00 &  0.10 &  0.32 &  0.33 \\
1270.00        &5.41  $\pm$ 0.47& 0.00 &  0.10 &  0.29 &  0.31 \\
1280.00        &5.50  $\pm$ 0.43& 0.00 &  0.10 &  0.30 &  0.31 \\
1290.00        &5.46  $\pm$ 0.42& 0.00 &  0.10 &  0.29 &  0.31 \\
1300.00        &5.13  $\pm$ 0.40& 0.00 &  0.07 &  0.28 &  0.29 \\
1310.00        &5.59  $\pm$ 0.52& 0.00 &  0.07 &  0.30 &  0.31 \\
1320.00        &5.44  $\pm$ 0.44& 0.00 &  0.09 &  0.29 &  0.31 \\
1330.00        &5.17  $\pm$ 0.38& 0.00 &  0.08 &  0.28 &  0.29 \\
1340.00        &4.70  $\pm$ 0.34& 0.00 &  0.08 &  0.25 &  0.27 \\
1350.00        &4.82  $\pm$ 0.41& 0.00 &  0.07 &  0.26 &  0.27 \\
1360.00        &4.68  $\pm$ 0.27& 0.00 &  0.09 &  0.25 &  0.27 \\
1370.00        &5.09  $\pm$ 0.43& 0.00 &  0.08 &  0.27 &  0.29 \\
1380.00        &4.21  $\pm$ 0.28& 0.00 &  0.09 &  0.23 &  0.25 \\ \hline
\end{tabular}
\end{center}
\end{table}

\begin{table}
\caption{The relative phase $\psi(s)$ of the amplitudes $A_{\omega\pi}$ and
         $A_{\rho\pi}$. }
\label{tab3}
\begin{center}
\begin{tabular}[t]{ccc}
$\sqrt[]{s}$(MeV)&$\psi(s)$(degree)&$P(\chi^2_0)$ \\ \hline
1100&  -57$\pm_{56}^{57}$&0.59 \\
1110&  -66$\pm_{74}^{66}$&0.57 \\
1120&   -1$\pm_{55}^{43}$&0.23 \\
1130&   37$\pm$        42&0.33 \\
1140&  130$\pm_{35}^{33}$&0.32 \\
1150&   60$\pm$       180&0.86 \\
1160&  -10$\pm_{39}^{35}$&0.03 \\
1180&   25$\pm_{28}^{30}$&0.98 \\
1190&  -20$\pm_{60}^{53}$&0.28 \\ 
1200&   23$\pm_{33}^{32}$&0.79 \\
1210&  131$\pm_{47}^{45}$&0.48 \\ 
1220&  -16$\pm$51        &0.44 \\
1230& -102$\pm$37        &0.28 \\
1240&  -21$\pm_{68}^{45}$&0.45 \\
1250&   26$\pm_{40}^{39}$&0.46 \\
1260&  -14$\pm_{61}^{48}$&0.18 \\
1270&  -26$\pm_{67}^{45}$&0.12 \\
1280&    1$\pm_{36}^{31}$&0.79 \\
1290&   23$\pm_{49}^{46}$&0.67 \\
1300&  -17$\pm_{36}^{33}$&0.33 \\
1310&   32$\pm_{34}^{33}$&0.42 \\
1320&  -34$\pm_{67}^{54}$&0.25 \\
1330&   41$\pm_{28}^{28}$&0.32 \\ 
1340&   30$\pm_{26}^{25}$&0.52 \\
1350&   49$\pm_{39}^{37}$&0.82 \\
1360&   35$\pm_{24}^{23}$&0.02 \\
1370&   19$\pm_{51}^{43}$&0.17 \\
1380&   23$\pm_{37}^{33}$&0.88 \\ \hline
\end{tabular} 
\end{center}
\end{table}

\begin{table}[h]
\caption{The results of the fit, taking into account three $\omega^i$
         resonances.}
\label{tab44}
\begin{center}
\begin{tabular}[t]{cccccc}
 $i$&$m_{\omega^i}$, MeV&$\Gamma_{i}$, MeV&$\sigma(\omega^i\to 3\pi)$, nb&
 $\sigma(\omega^i\to \omega\pi^+\pi^-)$, nb&$\phi_{\omega\omega^i}$ \\ \hline
 1&$1249 \pm^{42}_{87}$&$404 \pm^{88}_{81}$&
 $0.22 \pm^{0.23}_{0.17}$&--&$180^\circ$ \\
 2&$1428 \pm^{64}_{52}$&$765 \pm^{395}_{272}$&$2.02 \pm^{0.50}_{0.58}$&
 $0.05 \pm^{0.06}_{0.04}$&$180^\circ$ \\
 3&$1773 \pm^{30}_{26}$&$483 \pm^{93}_{73}$&$2.43 \pm^{0.56}_{0.47}$&
 $2.50 \pm^{0.33}_{0.31}$&$0^\circ$. \\ \hline
\end{tabular} 
\end{center}
\end{table}

\begin{table}
\caption{Fit results for the $e^+e^-\to\pi^+\pi^-\pi^0$ and $\omega\pi^+\pi^-$
         cross sections. The column number $N$ corresponds to the different
	 models for $A_{\omega\pi}$ amplitude.
	 $N_{3\pi}^{(SND)}$, $N_{3\pi}^{(DM2)}$ and $N_{\omega\pi\pi}^{(DM2)}$
	 is the number of
	 fitted points of the processes $e^+e^-\to\pi^+\pi^-\pi^0$ and
         $\omega\pi^+\pi^-$ obtained in SND and DM2 experiments. The DM2 data
	 was used in the fit as published in Ref.\cite{dm2}.}
\label{tab4}
\begin{center}
\begin{tabular}[t]{lccc}
 $N$&1&2&3 \\ \hline
$\sigma(\phi\to 3\pi)$, nb&647$\pm$4&646$\pm$4&647$\pm$4 \\
$m_{\omega^\prime}$, MeV&
1506$\pm^{40}_{32}$&1465$\pm^{33}_{38}$&1481$\pm^{35}_{30}$\\
$\Gamma_{\omega^\prime}$, MeV&
1322$\pm^{274}_{202}$&1037$\pm^{202}_{153}$&1079$\pm^{202}_{160}$\\
$\sigma(\omega^\prime\to 3\pi)$, nb&
3.31$\pm 0.49$&3.44$\pm^{0.46}_{0.47}$&3.56$\pm^{0.43}_{0.44}$\\
$\sigma(\omega^\prime\to \omega\pi^+\pi^-)$, nb&
0.03$\pm^{0.08}_{0.03}$&0.03$\pm^{0.07}_{0.03}$&0.03$\pm^{0.09}_{0.03}$\\
$\phi_{\omega\omega^\prime}$&$180^\circ$&$180^\circ$&$180^\circ$ \\
$m_{\omega^{\prime\prime}}$, MeV&
1798$\pm^{43}_{34}$&1801$\pm^{43}_{33}$&1793$\pm^{41}_{33}$\\
$\Gamma_{\omega^{\prime\prime}}$, MeV&
581$\pm^{176}_{119}$&580$\pm^{172}_{117}$&560$\pm^{162}_{120}$\\
$\sigma(\omega^{\prime\prime}\to 3\pi)$, nb&
1.72$\pm^{0.45}_{0.40}$&1.27$\pm^{0.33}_{0.32}$&1.54$\pm^{0.40}_{0.35}$\\
$\sigma(\omega^{\prime\prime}\to \omega\pi^+\pi^-)$, nb&
1.51$\pm^{0.34}_{0.30}$&1.48$\pm^{0.33}_{0.30}$&1.53$\pm^{0.34}_{0.31}$\\
$\phi_{\omega\omega^{\prime\prime}}$&$0^\circ$&$0^\circ$&$0^\circ$ \\
$\chi^2_{3\pi(SND)}/N_{3\pi}^{(SND)}$&55.3/67&52.4/67&52.7/67\\
$\chi^2_{3\pi(DM2)}/N_{3\pi}^{(DM2)}$&40.2/18&42.8/18&39.5/18\\
$\chi^2_{\omega\pi\pi(DM2)}/N_{\omega\pi\pi}^{(DM2)}$&9.3/18&9.8/18&9.3/18\\ \hline
\end{tabular} 
\end{center}
\end{table}

\begin{table}
\caption{Fit results for the $e^+e^-\to\pi^+\pi^-\pi^0$ and $\omega\pi^+\pi^-$
         cross sections. The column number $N$ corresponds to the different
	 models for $A_{\omega\pi}$ amplitude. $N_{3\pi}^{(SND)}$,
	 $N_{3\pi}^{(DM2)}$ and $N_{\omega\pi\pi}^{(DM2)}$ is the number of
	 fitted points of the processes $e^+e^-\to\pi^+\pi^-\pi^0$ and
	 $\omega\pi^+\pi^-$ obtained in SND and DM2 experiments.
	 The DM2 data was increase by a factor 1.54.}
\label{tab5}
\begin{center}
\begin{tabular}[t]{lccc}
 $N$&1&2&3 \\ \hline
$\sigma(\phi\to 3\pi)$, nb&646$\pm$4&646$\pm$4&646$\pm$4 \\
$m_{\omega^\prime}$, MeV&
1513$\pm^{45}_{37}$&1472$\pm^{40}_{32}$&1491$\pm^{44}_{37}$\\
$\Gamma_{\omega^\prime}$, MeV&
1383$\pm^{300}_{229}$&1095$\pm^{240}_{174}$&1156$\pm^{257}_{189}$\\
$\sigma(\omega^\prime\to 3\pi)$, nb&
3.45$\pm 0.50$&3.57$\pm^{0.47}_{0.51}$&3.65$\pm^{0.47}_{45}$\\
$\sigma(\omega^\prime\to \omega\pi^+\pi^-)$, nb&
0.03$\pm^{0.10}_{0.03}$&0.03$\pm^{0.11}_{0.03}$&0.04$\pm^{0.12}_{0.04}$\\
$\phi_{\omega\omega^\prime}$&$180^\circ$&$180^\circ$&$180^\circ$ \\
$m_{\omega^{\prime\prime}}$, MeV&
1784$\pm^{38}_{31}$&1784$\pm^{38}_{31}$&1780$\pm^{38}_{31}$\\
$\Gamma_{\omega^{\prime\prime}}$, MeV&
563$\pm^{156}_{110}$&550$\pm^{147}_{104}$&544$\pm^{146}_{104}$\\
$\sigma(\omega^{\prime\prime}\to 3\pi)$, nb&
2.80$\pm^{0.67}_{0.58}$&2.29$\pm^{0.54}_{0.49}$&2.59$\pm^{0.61}_{0.52}$\\
$\sigma(\omega^{\prime\prime}\to \omega\pi^+\pi^-)$, nb&
2.35$\pm^{0.48}_{0.44}$&2.34$\pm^{0.48}_{0.41}$&2.40$\pm^{0.49}_{0.44}$\\
$\phi_{\omega\omega^{\prime\prime}}$&$0^\circ$&$0^\circ$&$0^\circ$ \\
$\chi^2_{3\pi(SND)}/N_{3\pi}^{(SND)}$&51.8/67&49.2/67&49.6/67\\
$\chi^2_{3\pi(DM2)}/N_{3\pi}^{(DM2)}$&22.1/18&22.7/18&22.1/18\\
$\chi^2_{\omega\pi\pi(DM2)}/N_{\omega\pi\pi}^{(DM2)}$&9.3/18&9.4/18&9.3
\end{tabular} 
\end{center}
\end{table}

\begin{center}
\begin{figure}
\epsfig{figure=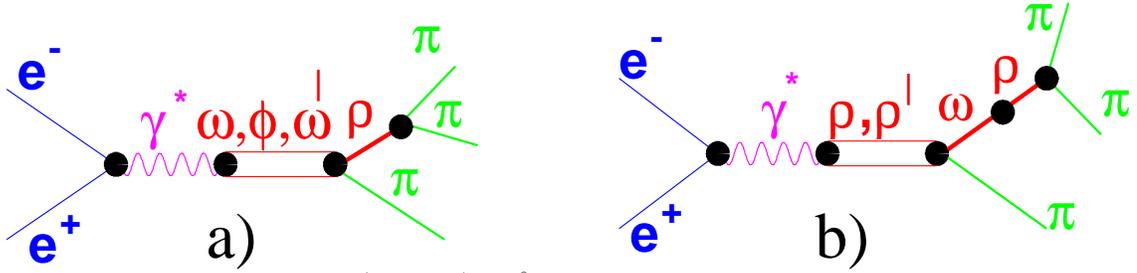,width=15cm}
\caption{The $e^+e^-\to\pi^+\pi^-\pi^0$ transition diagrams.}
\label{diag}
\end{figure}
\begin{figure}
\epsfig{figure=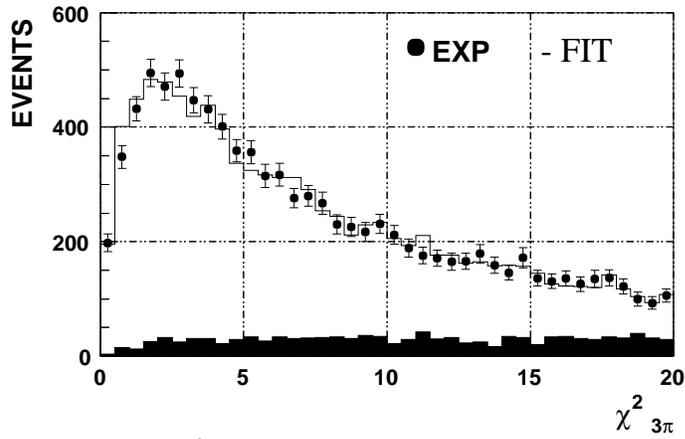,width=9cm}
\caption{The experimental $\chi^2_{3\pi}$ distribution, fitted by sum of
         distributions for signal and background. The background contribution
         is shown by filled histogram.}
\label{xi2u}
\end{figure}
\begin{figure}
\epsfig{figure=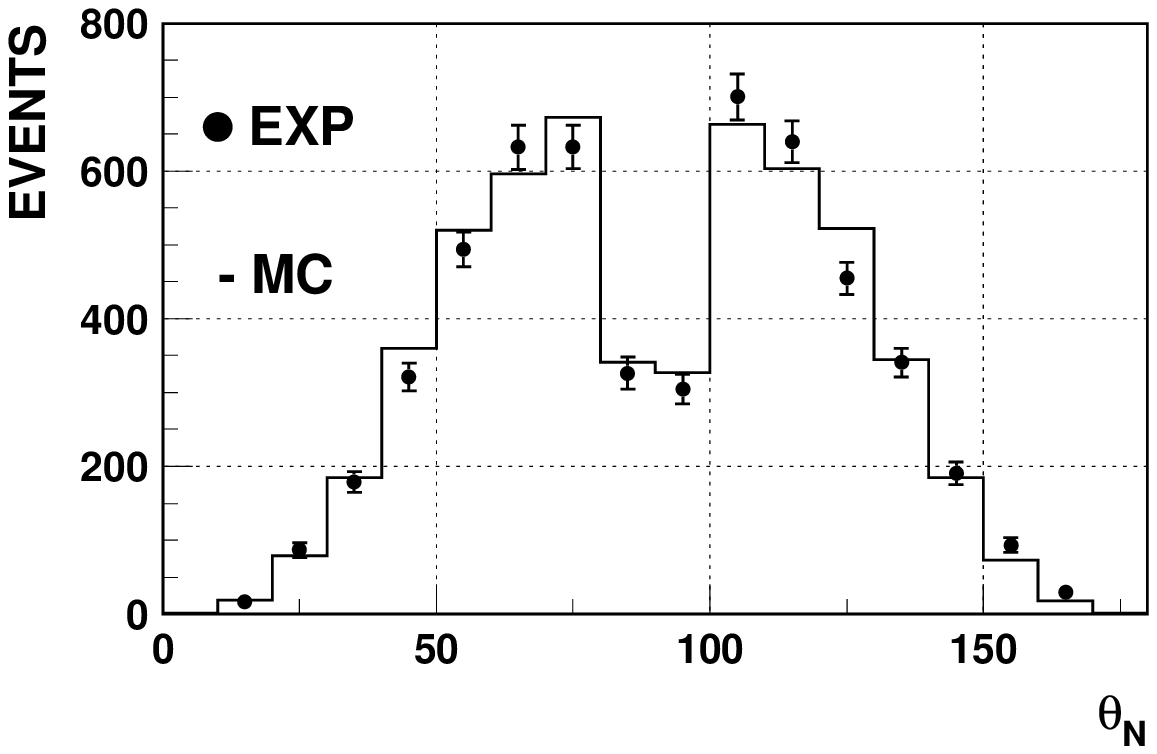,width=9cm}
\caption{The angle between the normal to the production plane and $e^+e^-$
         beam direction for $e^+e^- \to \pi^+\pi^-\pi^0$ events.}
\label{tetn}
\epsfig{file=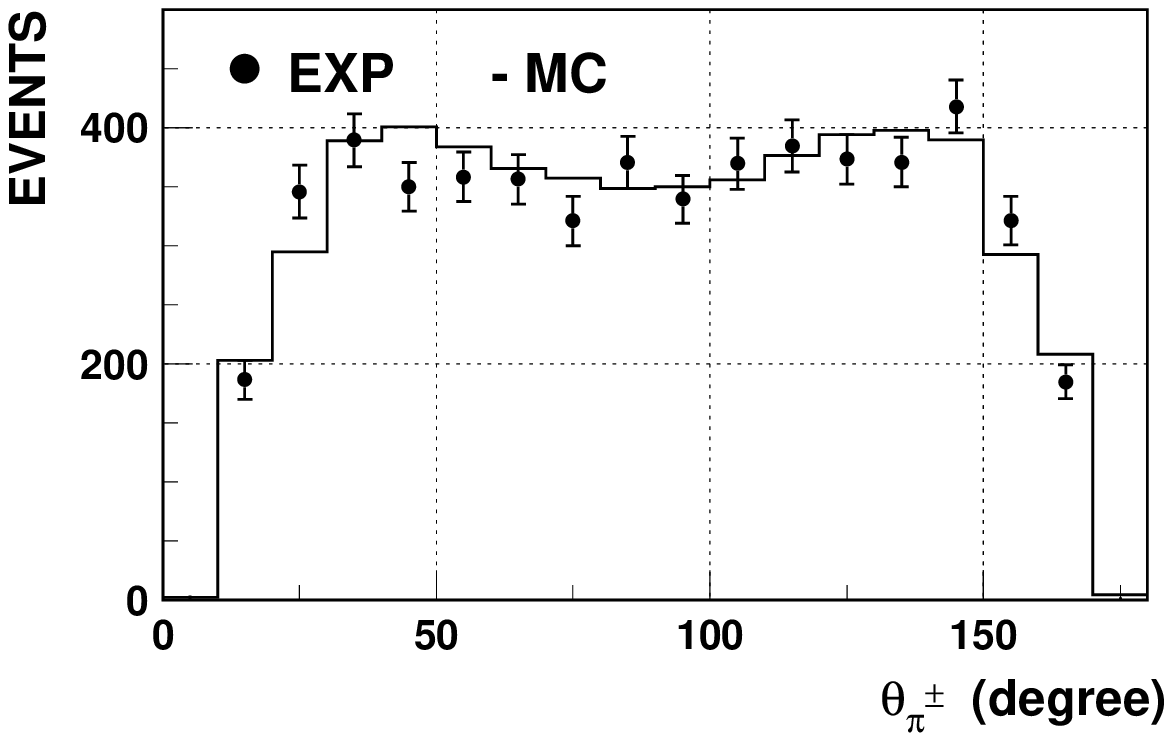,width=9cm}
\caption{The $\theta$ distribution of charged
pions from the reaction $e^+e^- \to \pi^+\pi^-\pi^0$.}
\label{teu12}
\epsfig{file=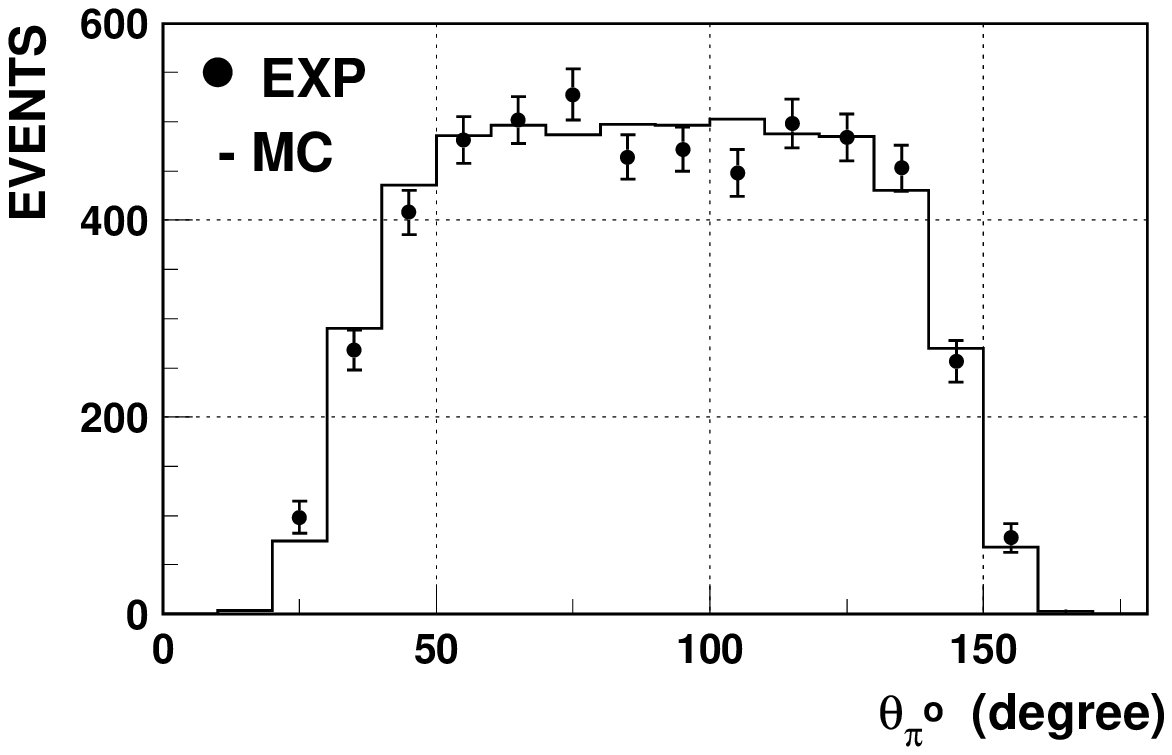,width=9cm}
\caption{The $\theta$ distribution of neutral pions from the reaction
$e^+e^- \to \pi^+\pi^-\pi^0$.}
\label{teu3}
\epsfig{file=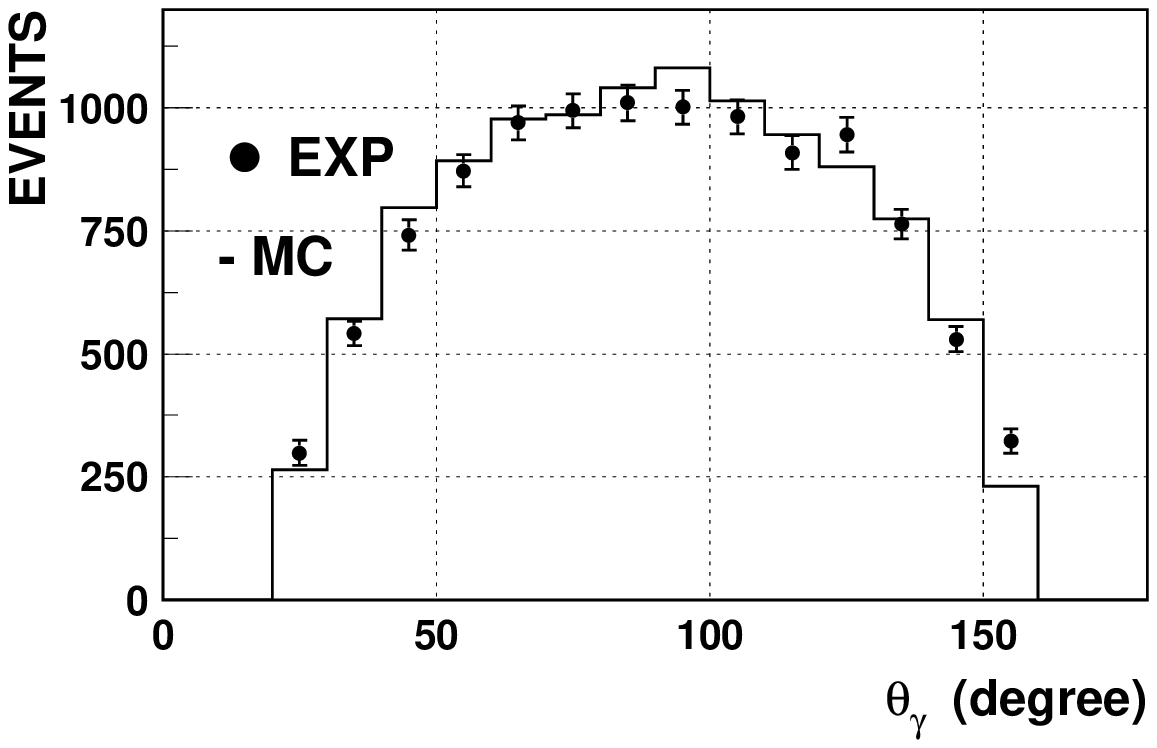,width=9cm}
\caption{The photon angular distribution.}
\label{teu45}
\epsfig{file=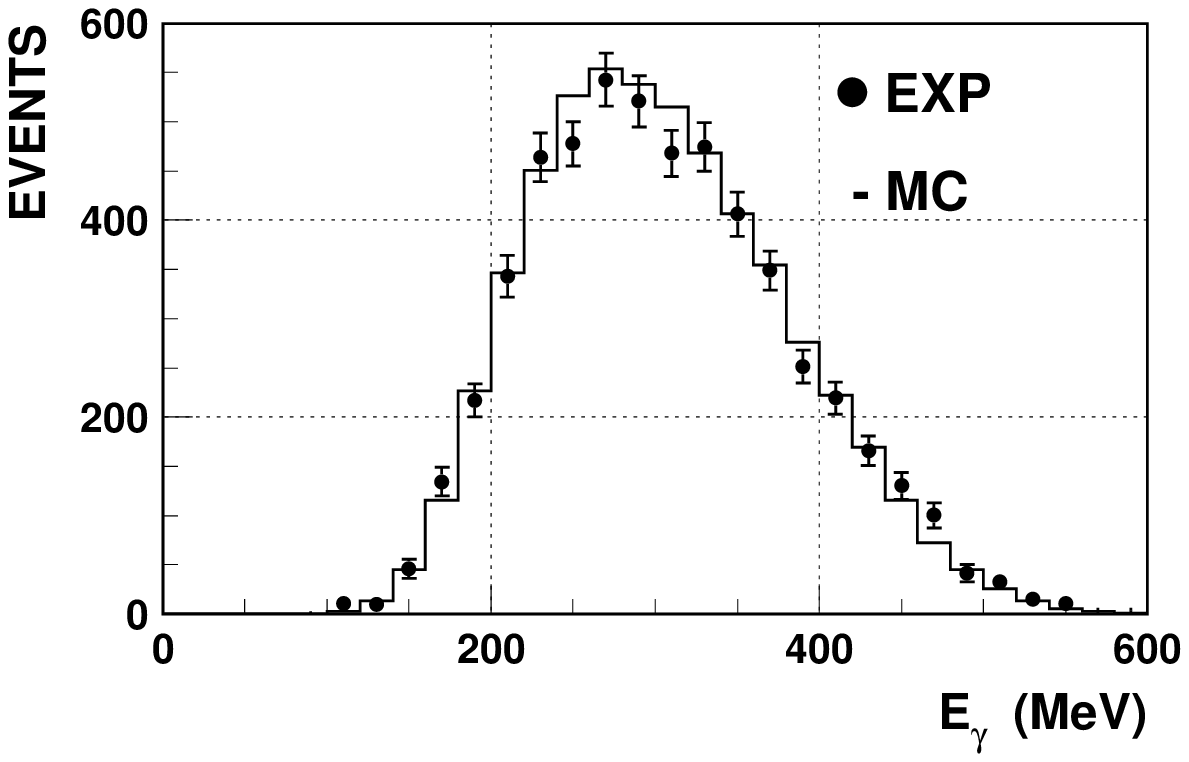,,width=9cm}
\caption{The energy distribution for the most energetic photon.}
\label{epu4}
\epsfig{file=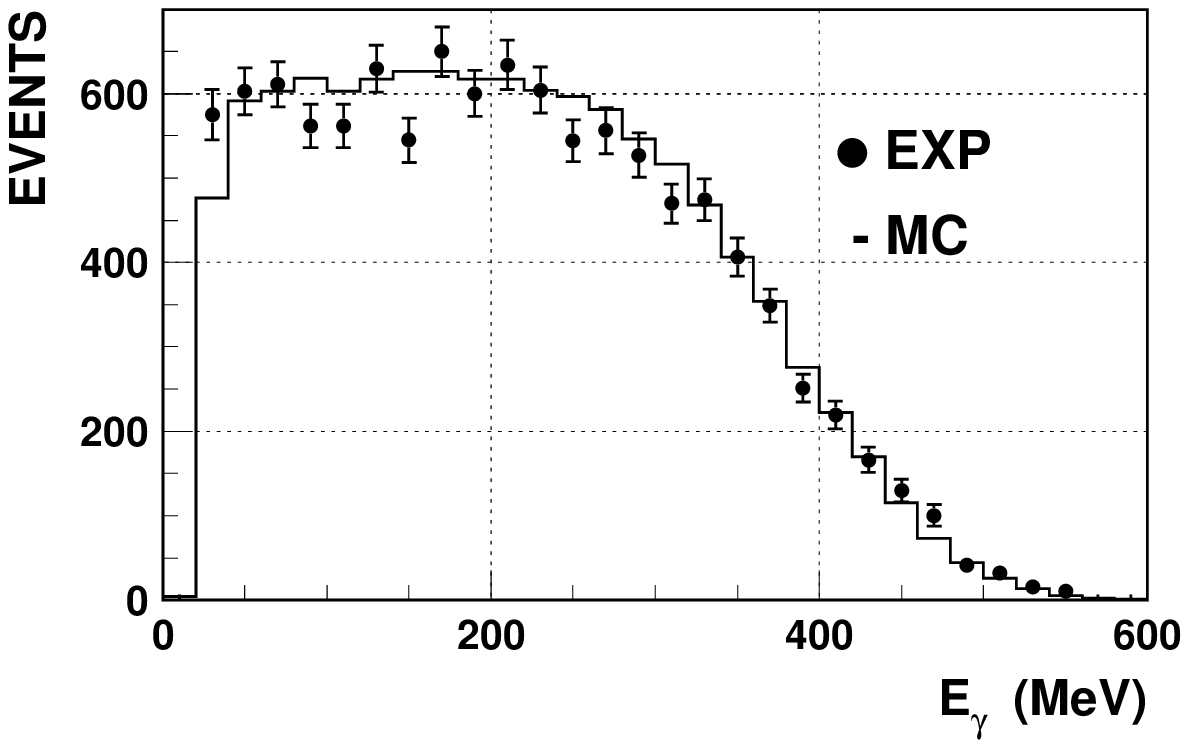,width=9cm}
\caption{Photon energy distribution.}
\label{epu45}
\epsfig{file=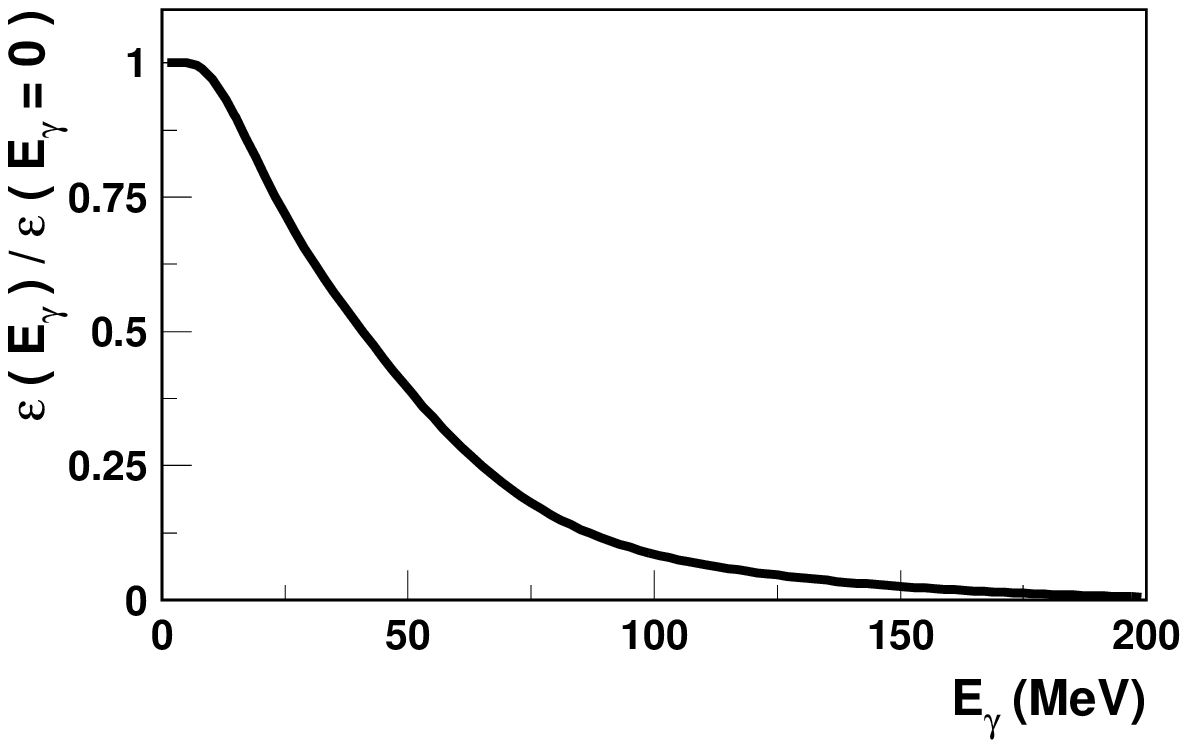,width=9cm}
\caption{The detection efficiency $\epsilon(E_\gamma)$ dependence on the
radiated photon energy $E_\gamma$ for $e^+e^-\to\pi^+\pi^-\pi^0(\gamma)$
events at $\sqrt[]{s}=1200$ MeV, obtained by simulation.}
\label{efrad}
\epsfig{file=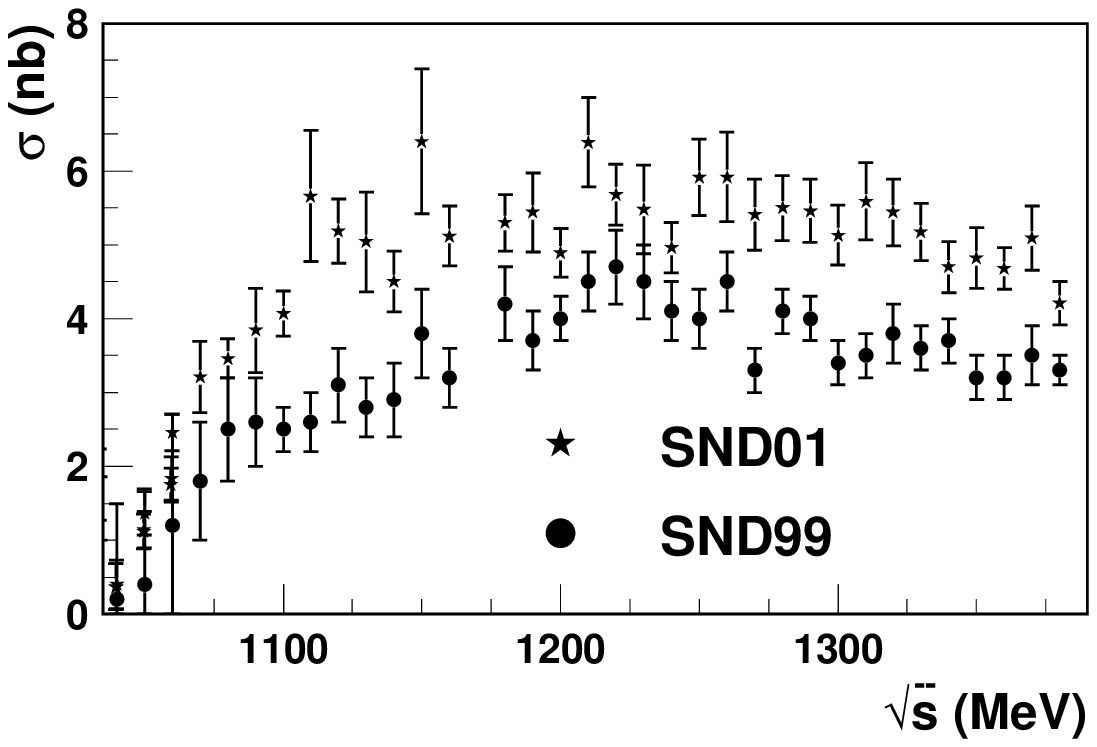,width=9cm}
\caption{Comparison of the  $e^+e^-\to\pi^+\pi^-\pi^0$ cross section obtained
in previous SND work \cite{sndmhad} (dots) and in this one (stars).}
\label{sis}
\epsfig{file=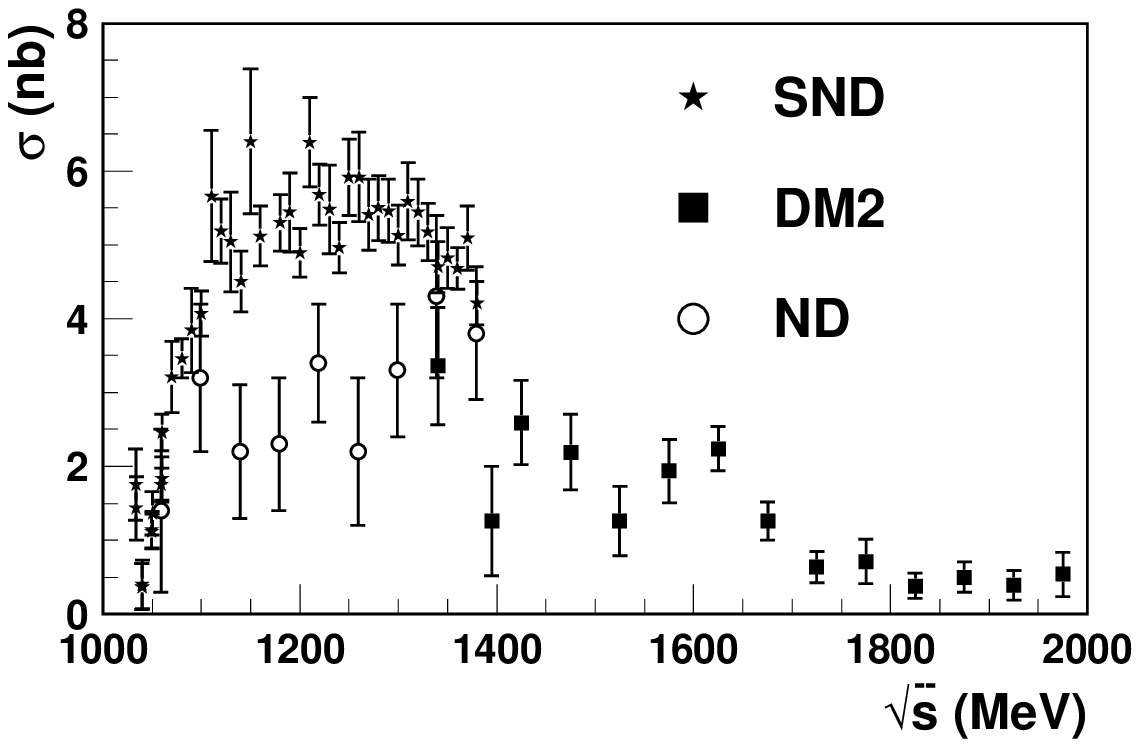,width=9cm}
\caption{The $e^+e^-\to\pi^+\pi^-\pi^0$ cross section at $\sqrt[]{s}$ from
         1030 to 2000 MeV. SND, ND \cite{nd} and DM2 \cite{dm2} results are
         shown.}
\label{cs}
\epsfig{file=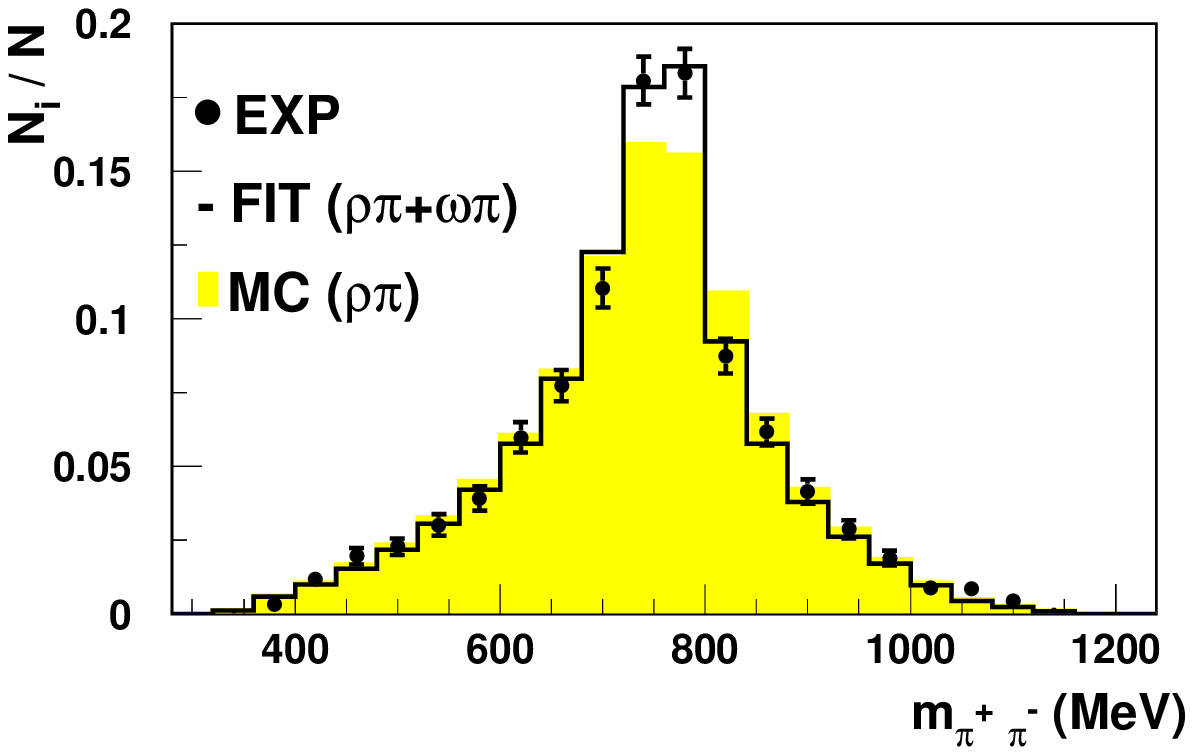,width=9cm}
\caption{The $\pi^+\pi^-$ invariant mass spectrum
at $\sqrt[]{s}$ from 1200 to 1380 MeV.}
\label{neu}
\epsfig{file=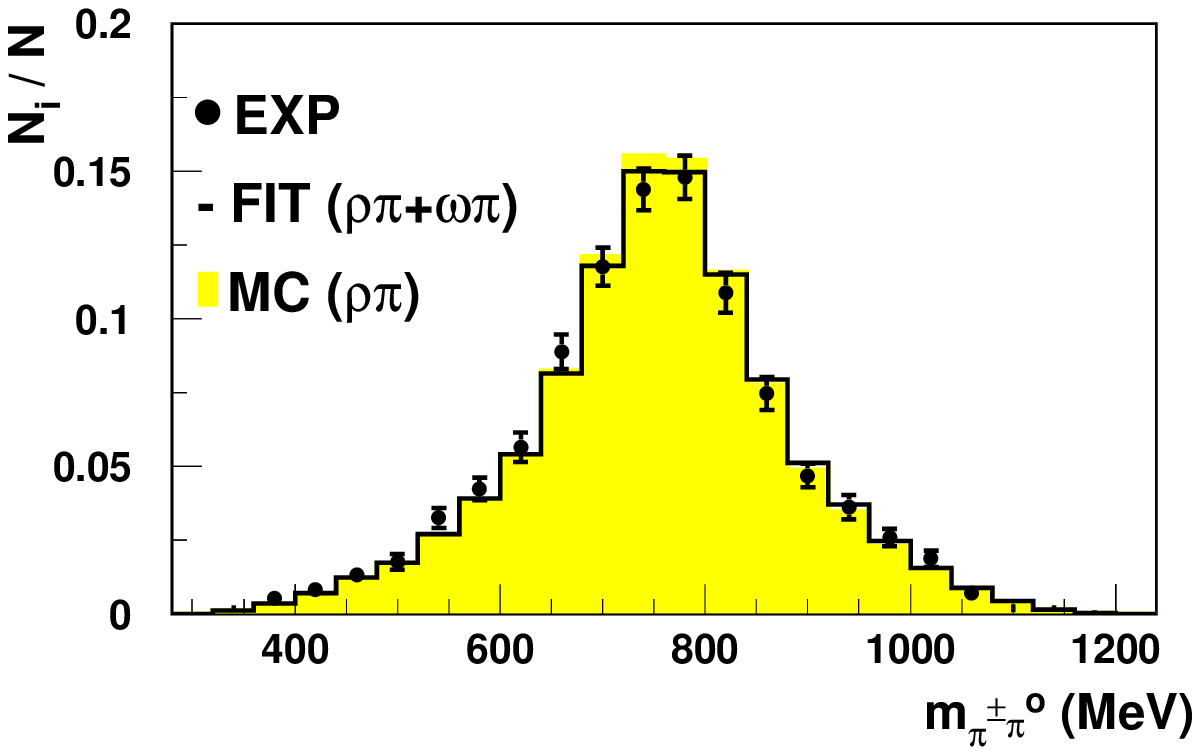,width=9cm}
\caption{The $\pi^\pm\pi^0$ invariant mass spectrum at $\sqrt[]{s}$ from 1200
to 1380 MeV.}
\label{cha}
\epsfig{file=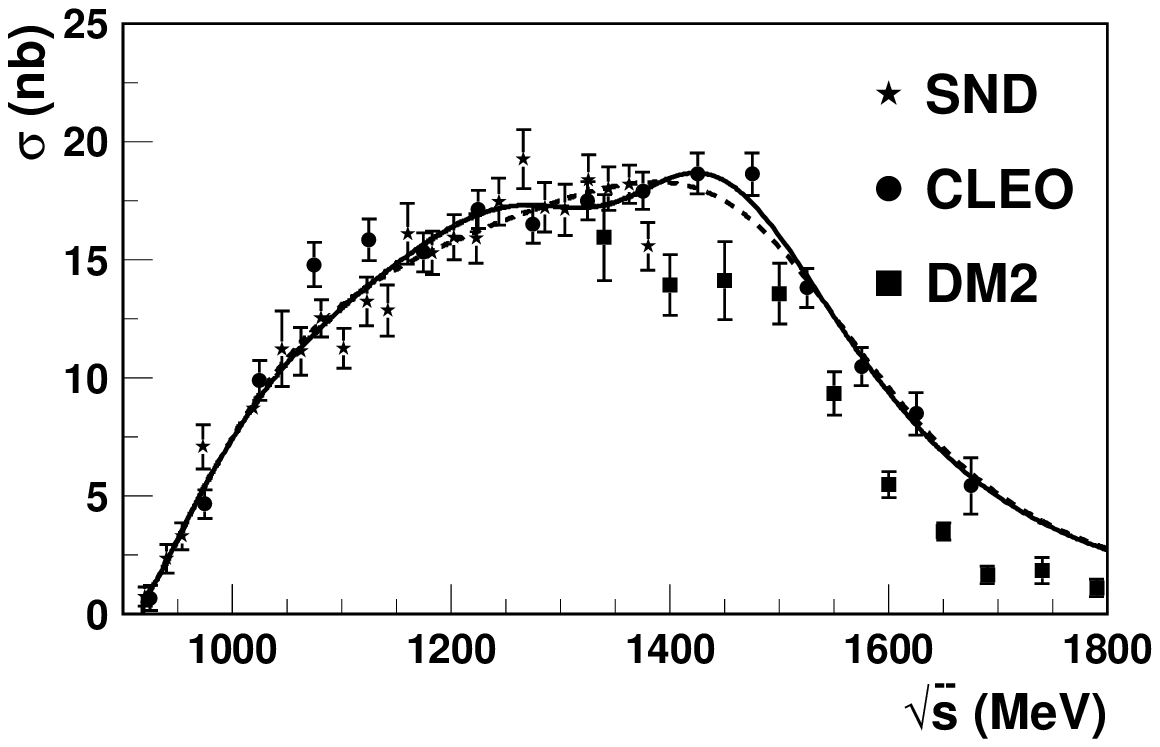,width=9cm}
\caption{The $e^+e^-\to\omega\pi$ cross section.
 The SND \cite{ppg}, CLEO2 \cite{cleo4p} and DM2 \cite{dm2omp} data are shown.
 The solid curve is the cross section energy dependence in the case when
 $A_{\omega\pi}=A_{\rho\to\omega\pi}+A_{\rho^\prime\to\omega\pi}+
 A_{\rho^{\prime\prime}\to\omega\pi}$, dashed curve is energy dependence 
 in the case
 $A_{\omega\pi}=A_{\rho\to\omega\pi}+A_{\rho^{\prime\prime}\to\omega\pi}$.}
\label{omp}
\epsfig{file=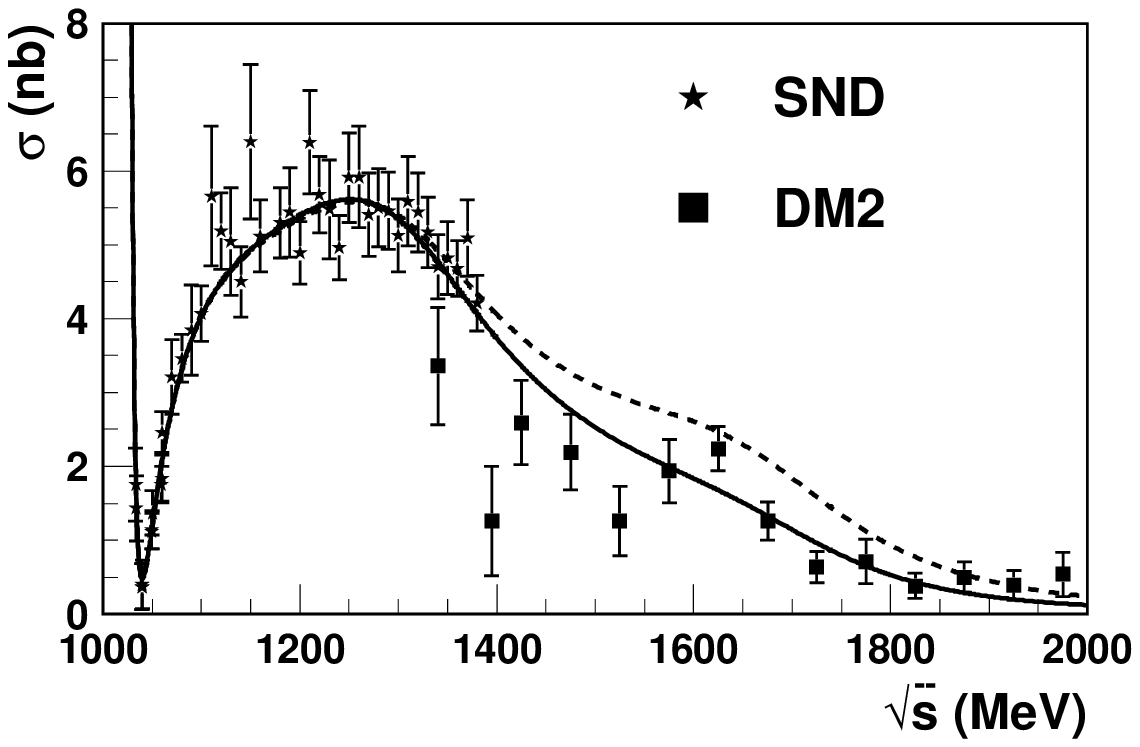,,width=9cm}
\caption{The $e^+e^-\to\pi^+\pi^-\pi^0$ cross section. SND and  DM2 \cite{dm2}
data are shown. Curves are the fits results. Dashed curve is the fit
with DM2 data increase by a factor 1.54.}
\label{cs3}
\epsfig{file=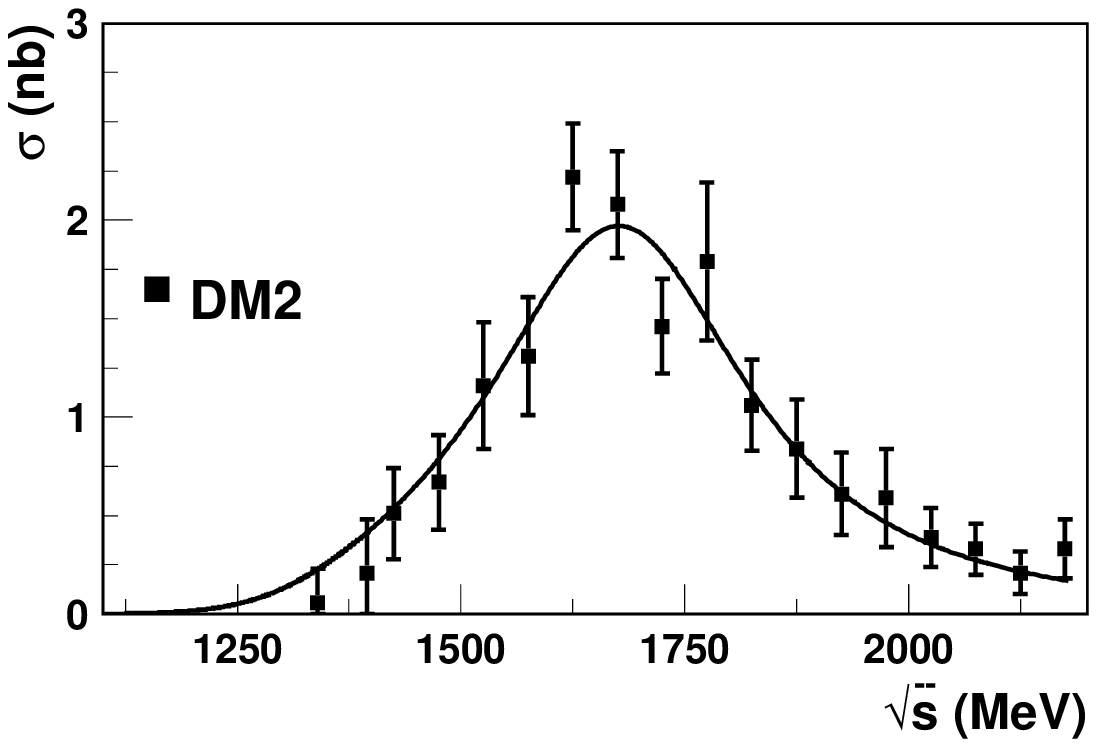,width=9cm}
\caption{The $e^+e^-\to\omega\pi^+\pi^-$ cross
section.  DM2 \cite{dm2} data are shown; curve is the fit result.}
\label{cs4}
\end{figure}
\begin{figure}
\epsfig{file=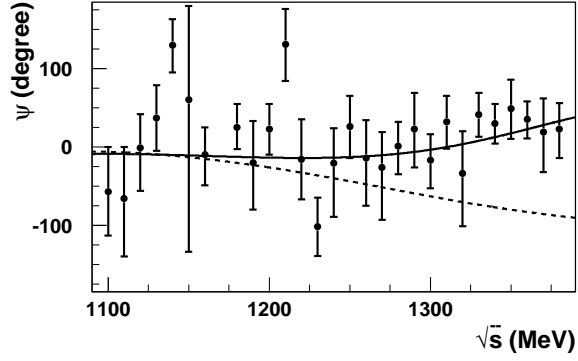,width=9cm}
\caption{The comparison of the relative phase $\psi(s)$ of the amplitudes
$A_{\rho\pi}$ and $A_{\omega\pi}$ measured in this work (dots) with theoretical
dependences: solid curve is the phase $\psi(s)$ in the case
$A_{\omega\pi}=A_{\rho\to\omega\pi}+A_{\rho^\prime\to\omega\pi}+
A_{\rho^{\prime\prime}\to\omega\pi}$, dashed curve is the phase $\psi(s)$ in 
the case
$A_{\omega\pi}=A_{\rho\to\omega\pi}+A_{\rho^{\prime\prime}\to\omega\pi}$}
\label{faz2}
\end{figure}
\end{center}
\end{document}